\RequirePackage{lineno}
\documentclass[12pt,tightenlines,superscriptaddress,a4paper]{revtex4-2}

\usepackage[page,title]{appendix}
\usepackage{url}
\usepackage{placeins}
\usepackage{amsfonts}
\usepackage[dvipsnames]{xcolor}
\usepackage{color}
\usepackage{xspace}
\usepackage{tikz}
\usepackage[utf8]{inputenc}										
\usepackage{gensymb}

\usepackage{subfigure}

\usepackage{svg}

\DeclareUnicodeCharacter{2212}{-}
\DeclareUnicodeCharacter{03B8}{$\theta$}

\usetikzlibrary{arrows}
\usetikzlibrary{shapes}
\usetikzlibrary{trees}
\usetikzlibrary{matrix}
\usetikzlibrary{arrows} 			
\renewcommand{\draft}[1]{{\color{blue}#1}}

\usepackage{graphicx,psfrag,bm}
\usepackage{mathrsfs,amsfonts}
\usepackage{epsfig,cancel}
\usepackage{epstopdf}
\usepackage{mciteplus}
\usepackage{latexsym}
\usepackage{amsthm}
\usepackage{amsmath}
\usepackage{amssymb}
\usepackage{hepunits}
\usepackage{hyperref}
\usepackage{bbm}
\usepackage{xfrac}
\usepackage{colordvi}
\usepackage{comment}
\usepackage{dcolumn}
\usepackage{times,wrapfig}
\usepackage{slashed}
\usepackage{booktabs}
\usepackage{relsize}

\hypersetup{pdftitle={Photon-rejection Power of the Light Dark Matter eXperiment in an 8 GeV Beam}}

\newcommand{\be}{\begin{eqnarray}}
\newcommand{\ee}{\end{eqnarray}}

\newcommand{\benum}{\begin{enumerate}}
\newcommand{\eenum}{\end{enumerate}}
\newcommand{\bi}{\begin{itemize}}
\newcommand{\ei}{\end{itemize}}
\newcommand{\geant}{\textsc{Geant4}\xspace}

\newcommand{\ecal}{ECal\xspace}

\newcommand{\pn}{photo-nuclear\xspace}

\def\babar{\mbox{\slshape B\kern-0.1em{\smaller A}\kern-0.1emB\kern-0.1em{\smaller A\kern-0.2em R}}}

\newcommand{\ednote}[1]{} 
\newcommand{\people}[1]{} 
\newcommand{\morepeople}[1]{} 

\colorlet{RED}{red}
\colorlet{BLUE}{blue}
\colorlet{ORANGE}{orange}

\begin{document}

%
%
\title{Photon-rejection Power of the Light Dark Matter eXperiment in an 8 GeV Beam}

\date{September 4, 2023}
\hspace*{0pt}\hfill 
{\small FERMILAB-PUB-23-433-PPD-T, \,SLAC-PUB-17550}

\bigskip
\author{Torsten~Åkesson}
\affiliation{Lund University, Department of Physics, Box 118, 221 00 Lund, Sweden}

\author{Cameron~Bravo}
\affiliation{SLAC National Accelerator Laboratory, Menlo Park, CA 94025, USA}

\author{Liam~Brennan}
\affiliation{University of California at Santa Barbara, Santa Barbara, CA 93106, USA}

\author{Lene~Kristian~Bryngemark}
\affiliation{Lund University, Department of Physics, Box 118, 221 00 Lund, Sweden}

\author{Pierfrancesco~Butti}
\affiliation{SLAC National Accelerator Laboratory, Menlo Park, CA 94025, USA}

\author{E.~Craig~Dukes}
\affiliation{University of Virginia, Charlottesville, VA 22904, USA}

\author{Valentina~Dutta}
\affiliation{Carnegie Mellon University, Pittsburgh, PA 15213, USA}

\author{Bertrand~Echenard}
\affiliation{California Institute of Technology, Pasadena, CA 91125, USA}

\author{Thomas~Eichlersmith}
\affiliation{University of Minnesota, Minneapolis, MN 55455, USA}

\author{Jonathan~Eisch}
\affiliation{Fermi National Accelerator Laboratory, Batavia, IL 60510, USA}

\author{Einar~El\'{e}n}
\affiliation{Lund University, Department of Physics, Box 118, 221 00 Lund, Sweden}

\author{Ralf~Ehrlich}
\affiliation{University of Virginia, Charlottesville, VA 22904, USA}

\author{Cooper~Froemming}
\affiliation{University of Minnesota, Minneapolis, MN 55455, USA}

\author{Andrew~Furmanski}
\affiliation{University of Minnesota, Minneapolis, MN 55455, USA}

\author{Niramay~Gogate}
\affiliation{Texas Tech University, Lubbock, TX 79409, USA}

\author{Chiara~Grieco}
\affiliation{University of California at Santa Barbara, Santa Barbara, CA 93106, USA}

\author{Craig~Group}
\affiliation{University of Virginia, Charlottesville, VA 22904, USA}

\author{Hannah~Herde}
\affiliation{Lund University, Department of Physics, Box 118, 221 00 Lund, Sweden}

\author{Christian~Herwig}
\affiliation{Fermi National Accelerator Laboratory, Batavia, IL 60510, USA}

\author{David~G.~Hitlin}
\affiliation{California Institute of Technology, Pasadena, CA 91125, USA}

\author{Tyler~Horoho}
\affiliation{University of Virginia, Charlottesville, VA 22904, USA}

\author{Joseph~Incandela}
\affiliation{University of California at Santa Barbara, Santa Barbara, CA 93106, USA}

\author{Wesley~Ketchum}
\affiliation{Fermi National Accelerator Laboratory, Batavia, IL 60510, USA}

\author{Gordan~Krnjaic}
\affiliation{Fermi National Accelerator Laboratory, Batavia, IL 60510, USA}

\author{Amina~Li}
\affiliation{University of California at Santa Barbara, Santa Barbara, CA 93106, USA}

\author{Jeremiah~Mans}
\affiliation{University of Minnesota, Minneapolis, MN 55455, USA}

\author{Phillip~Masterson}
\affiliation{University of California at Santa Barbara, Santa Barbara, CA 93106, USA}

\author{Sophie~Middleton}
\affiliation{California Institute of Technology, Pasadena, CA 91125, USA}

\author{Omar~Moreno}
\affiliation{SLAC National Accelerator Laboratory, Menlo Park, CA 94025, USA}

\author{Geoffrey~Mullier}
\affiliation{Lund University, Department of Physics, Box 118, 221 00 Lund, Sweden}

\author{Joseph~Muse}
\affiliation{University of Minnesota, Minneapolis, MN 55455, USA}

\author{Timothy~Nelson}
\affiliation{SLAC National Accelerator Laboratory, Menlo Park, CA 94025, USA}

\author{Rory~O'Dwyer}
\affiliation{Stanford University, Menlo Park, CA 94025, USA}

\author{Leo~\"{O}stman}
\affiliation{Lund University, Department of Physics, Box 118, 221 00 Lund, Sweden}

\author{James Oyang}
\affiliation{California Institute of Technology, Pasadena, CA 91125, USA}

\author{Jessica~Pascadlo}
\affiliation{University of Virginia, Charlottesville, VA 22904, USA}

\author{Ruth~P\"{o}ttgen}
\affiliation{Lund University, Department of Physics, Box 118, 221 00 Lund, Sweden}

\author{Luis~G.~Sarmiento}
\affiliation{Lund University, Department of Physics, Box 118, 221 00 Lund, Sweden}

\author{Philip~Schuster}
\affiliation{SLAC National Accelerator Laboratory, Menlo Park, CA 94025, USA}

\author{Matthew~Solt}
\affiliation{University of Virginia, Charlottesville, VA 22904, USA}

\author{Cristina~Mantilla~Suarez}
\affiliation{Fermi National Accelerator Laboratory, Batavia, IL 60510, USA}

\author{Lauren~Tompkins}
\affiliation{Stanford University, Menlo Park, CA 94025, USA}

\author{Natalia~Toro}
\affiliation{SLAC National Accelerator Laboratory, Menlo Park, CA 94025, USA}

\author{Nhan~Tran}
\affiliation{Fermi National Accelerator Laboratory, Batavia, IL 60510, USA}

\author{Erik~Wallin}
\affiliation{Lund University, Department of Physics, Box 118, 221 00 Lund, Sweden}

\author{Andrew~Whitbeck}
\affiliation{Texas Tech University, Lubbock, TX 79409, USA}

\author{Danyi~Zhang}
\affiliation{University of California at Santa Barbara, Santa Barbara, CA 93106, USA}

\begin{abstract}
The Light Dark Matter eXperiment (LDMX) is an electron-beam fixed-target experiment designed to achieve comprehensive model independent sensitivity to dark matter particles in the sub-GeV mass region.
An upgrade to the LCLS-II accelerator will increase the beam energy available to LDMX from 4 to 8\,GeV.
Using detailed GEANT4-based simulations, we investigate the effect of the increased beam energy on the capabilities to separate
signal and background, and demonstrate that the veto methodology developed for 4\,GeV successfully rejects photon-induced backgrounds for at least $2\times10^{14}$ electrons on target at 8\,GeV. 

\end{abstract}

\maketitle
\newpage
\tableofcontents
\newpage

\section{Introduction}
\label{sec:intro}
\parskip0pt

The Light Dark Matter eXperiment (LDMX) 
is a planned experiment designed primarily to search for dark matter particles with masses in the MeV to GeV region. It aims to observe the production of such particles in interactions of a primary electron beam with a thin target by measuring the energy loss and momentum change the electron undergoes in the target. All relevant background reactions are accompanied by the production of detectable Standard Model (SM) particles, albeit with low multiplicity and/or low energy in some cases. Thus, achieving the desired sensitivity depends on the efficiency with which these particles can be detected. 
The most challenging backgrounds arise from photo-nuclear (PN) reactions of hard bremsstrahlung photons, where `hard' is defined as carrying more than about three-quarters of the initial electron's energy. Rejecting these reactions is particularly difficult if their final state is only a single neutral hadron (mainly a neutron or $K_{L}^{0}$), or a single charged particle that decays in-flight with most of its energy being carried away by a neutrino. 

LDMX intends to take its first data with a low-intensity electron beam extracted from the Linac Coherent Light Source II (LCLS-II) at SLAC with an initial beam energy of 4\,GeV.  
The majority of the LDMX data, $10^{16}$ electrons on target (EoT) or more, will be collected at a beam energy of 8\,GeV after a corresponding upgrade of LCLS-II~\cite{Raubenheimer:2018wwc}.
A higher beam energy is expected to improve the sensitivity of LDMX for a number of reasons: (i) the signal yield
will increase \cite{WhitePaper}, 
(ii) the rates of several challenging PN event classes will decrease, and (iii) the PN reactions will typically have higher-multiplicity final states with more energetic particles, making them more likely to be detected. 
In this paper, we focus on aspect (iii) and quantify the energy increase's impact on the photon-induced background rejection, based on simulation studies. 
The baseline is the corresponding simulation studies done for a beam energy of 4\,GeV as reported in Ref.\,\cite{VetoPaper}. 
They show that photon-induced backgrounds at 4\,GeV beam energy can efficiently be rejected with the currently envisioned detector design, considering a sample of $4\times10^{14}$ EoT.

The paper is organised as follows: in Sec.~\ref{sec:ldmx} we review the LDMX detector design used in this study, as well as the signal and background characteristics. The simulation samples employed are summarised in Sec.~\ref{sec:sim}. Sec.~\ref{sec:energycomparison} shows comparisons of important variables at 4\,GeV and 8\,GeV.  In Sec.~\ref{sec:evtSel} we briefly recapitulate the selections applied at 4\,GeV, as well as their performance, followed by a description of the corresponding selections at 8\,GeV. 
The results are discussed in Sec.~\ref{sec:res}, and we conclude in Sec.~\ref{sec:concl}.

\section{The Light Dark Matter eXperiment}
\label{sec:ldmx}
A detailed description of the conceptual design of LDMX can be found in \cite{VetoPaper, WhitePaper}. Here, we briefly summarise the design-driving physics aspects and the chosen detector solutions, a GEANT4-based simulation~\cite{Agostinelli:2002hh,Allison:2016lfl} of which was used to obtain the results presented in this paper. 
While the discussion in the following sections is centered around the main motivation of searching for sub-GeV thermal relic dark matter, we emphasise that the resulting detector design has a much broader potential: quite generally, LDMX will enable measurements with electron beams in the forward region, using both invisible and visible signatures, as discussed for example in \cite{snowmass22,Berlin:2018bsc}.

\subsection{Light Dark Matter: Motivation and Signature}
\label{sec:thBkg}

A compelling hypothesis explaining the origin of dark matter (DM) is that DM is a thermal relic from the early Universe. Within this simple yet predictive formalism, the observed abundance of DM can be correctly reproduced for DM particle masses in the range of MeV to O(100) TeV, assuming some (small) non-gravitational interaction between DM and SM particles. For DM particles to be lighter than a few GeV, this interaction has to be mediated by a new, light particle, to avoid overproduction of DM compared to the relic abundance. The existence of such an interaction implies a DM production mechanism at accelerators, which LDMX aims to exploit. 

In this paper, we employ the widely-used benchmark of a kinetically-mixed dark photon ($A'$)~\cite{PhysRevLett.115.251301} as our signal model. 
Despite its simplicity, it captures the most critical features of a DM signal and is indicative of sensitivity to classes of models with richer dark sector structures.
It contains four parameters: the kinetic mixing parameter, $\varepsilon$, the coupling between the dark photon and DM, $\alpha_D$, and the masses of the DM particles, $m_{\chi}$, and the dark photon, $m_{A'}$. We consider the case  $m_{A'}>2m_{\chi}$, such that the dark photon will predominantly decay invisibly into dark matter.
The common benchmark of $m_{A'}=3m_{\chi}$ is used throughout this paper.
The production of dark matter in the target will proceed via bremsstrahlung of a dark photon and its subsequent decay into dark matter.
The kinematics are qualitatively similar for production through an off-shell mediator or a contact interaction.
In contrast to ordinary bremsstrahlung, the massive dark photon will carry more of the incoming electron's energy.
Momentum conservation then requires the outgoing electron to have a sizeable transverse momentum. The LDMX detector design aims to measure both the energy of the outgoing recoil electron, $E_{recoil}$, and its transverse momentum, $p_{T}$. The recoil energy, or equivalently the missing energy, $E_{miss}=E_{beam}-E_{recoil}$, is measured with an electromagnetic calorimeter (ECal) downstream of the target. 
Upstream of the target, as well as between the target and the ECal, tracking detectors in a magnetic field will measure the momentum difference before and after the electron has passed through the target. A hadronic calorimeter (HCal) surrounds the ECal as much as possible to catch products of SM reactions that would be invisible to the ECal. 
With this setup, the DM signature is an electron losing most of its energy and acquiring sizeable transverse momentum in an otherwise completely empty detector. 

\begin{figure}[t]
    \centering    
    \smaller
    \def\svgwidth{7in}
    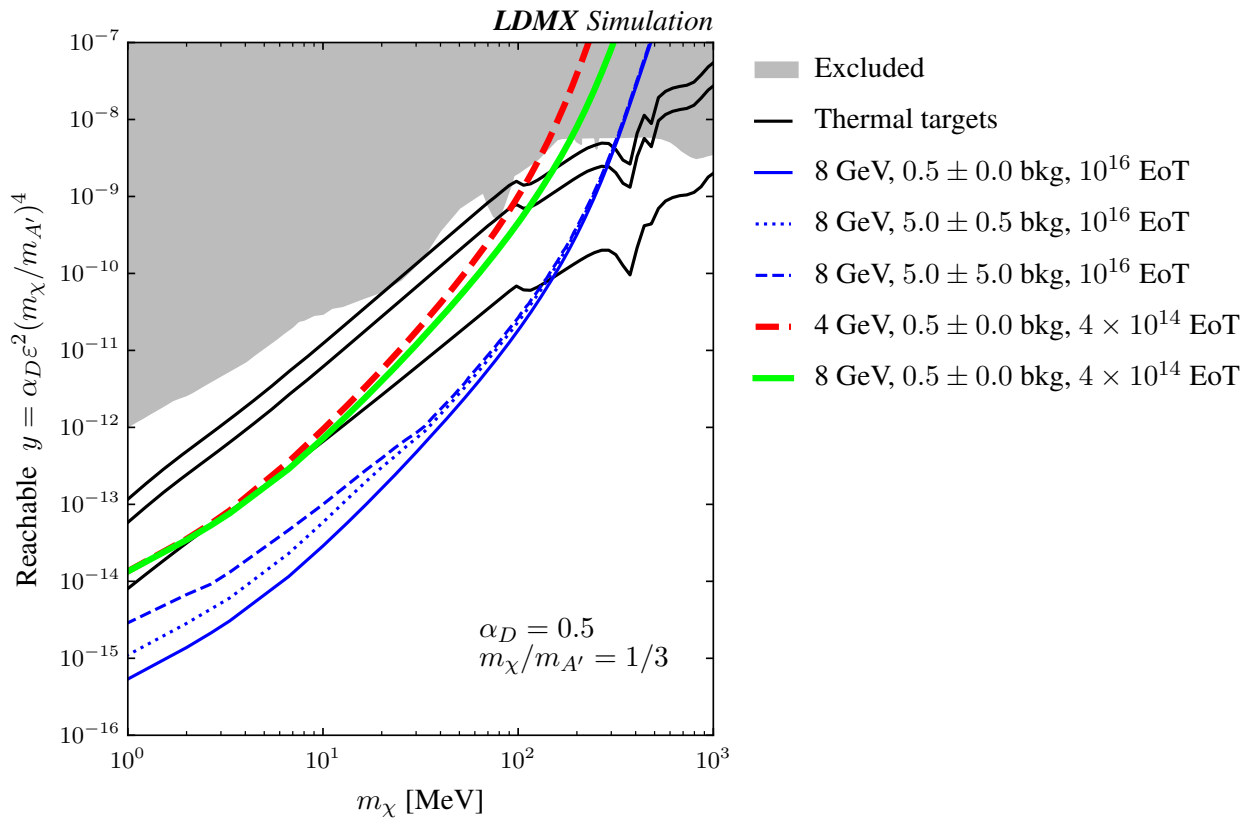
    \caption{The projected reach of LDMX in the dark bremsstrahlung model at 4 and 8 GeV beam energies. 
    Several scenarios are shown for the $10^{16}$ electrons on target (EoT) run at 8\,GeV assuming different background levels.
    Three thermal relic targets are shown as black lines, corresponding from top to bottom, to scalar, Majorana fermion, and Pseudo-Dirac fermion models of DM \cite{WhitePaper}. The region in grey is already excluded by other experiments \cite{deniverville:2016rqh,Lees:2017lec,Aguilar-Arevalo:2017mqx,NA64:2019imj,batell:2014mga,Andreev:2021fzd,COHERENT:2021pvd,CCM:2021leg}. A 50\% signal efficiency is assumed across the entire $m_\chi$ mass spectrum at both beam energies.  
    } 
    \label{fig:Reach8GeV}
\end{figure}

Fig.~\ref{fig:Reach8GeV} shows projections of the exclusion sensitivity of LDMX for different scenarios, assuming a flat 50\% signal efficiency across the $m_{A'}$ mass range. The sensitivity is shown in terms of the interaction strength parameter $y=\alpha_D \varepsilon^2 (m_\chi/m_{A'})^4$ and the DM particle mass, $m_{\chi}$. 
 The thermal targets, i.e. the $(y, m_{\chi})$ combinations that reproduce the measured abundance of dark matter as a thermal relic for different dark matter particle types, are not yet excluded by other experiments for these benchmark models. 
 The discrimination between signal events and remaining background events is further improved by recoil electron $p_{T}$ information, with the strongest improvement for higher $m_{\chi}$.
 The LDMX projections would exclude yet uncovered regions of the parameter space. 
 This figure considers only the reach of LDMX for dark bremsstrahlung production, while the sensitivity at higher $m_{\chi}$ may be improved by also considering invisible vector meson decays \cite{Schuster:2021mlr}.
 
 The projection for the 4\,GeV running period is shown, assuming that all background events can be vetoed. A corresponding dataset of the same size at 8\,GeV illustrates the gain purely from increasing the beam energy, which is most relevant at dark matter masses above 10 MeV, and continues to increase for larger masses. 
A data set of $10^{16}$\,EoT collected at 8\,GeV  would significantly extend the reach over the entire mass range.
Variations of the number of non-rejected background events or its uncertainty affect the potential reach, emphasising the need to minimise both the expected rate of such false-positive background events and its uncertainty. 

\subsection{Detector Components} 
\begin{figure}
    \centering
    \includegraphics[width=0.95\textwidth]{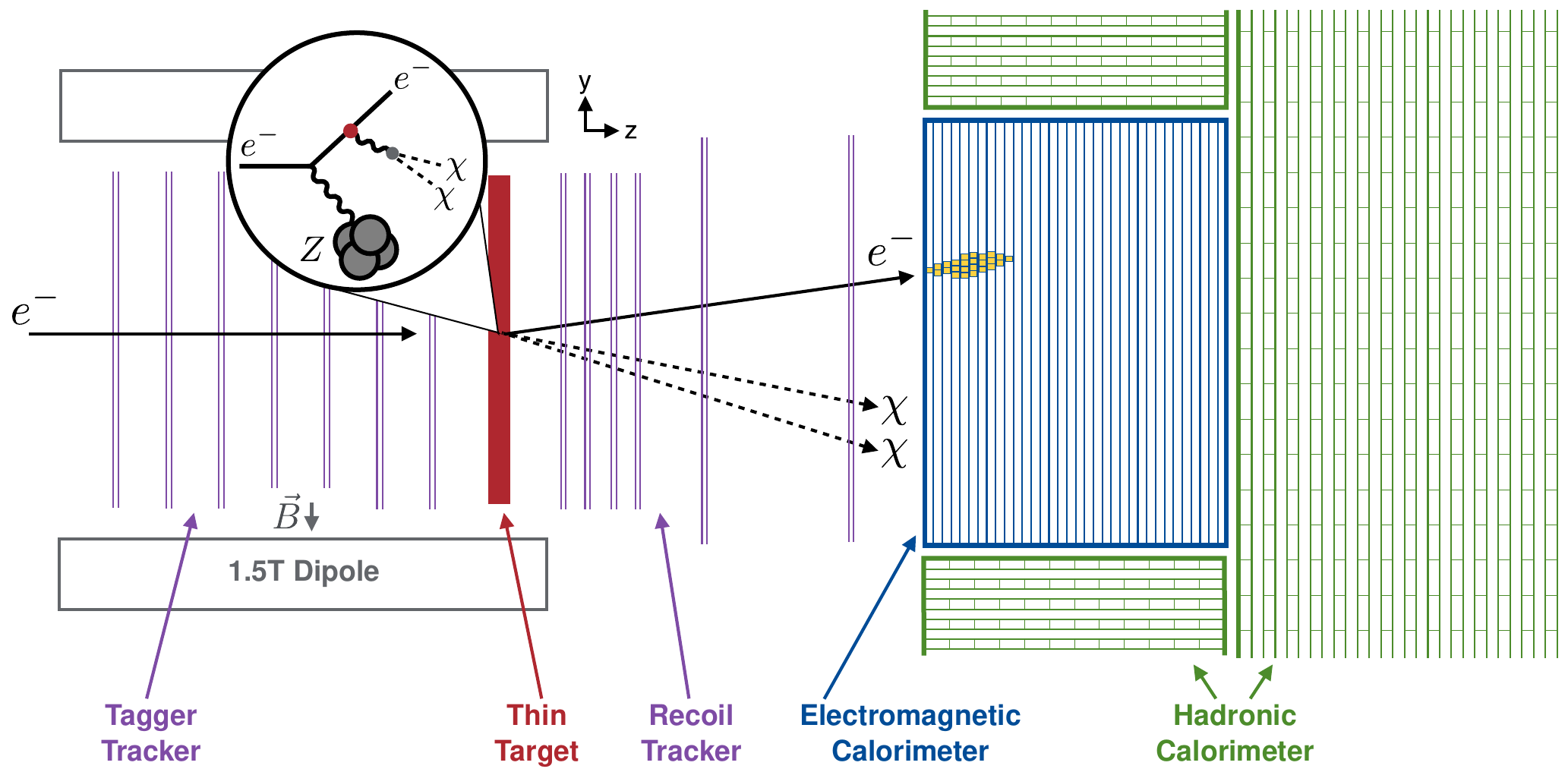}
    \caption{ A diagram of the LDMX detector apparatus, illustrating production of DM in the target from a scattering beam electron, and the corresponding response of the various sub-systems to the missing-momentum signature.  The drawing is not to scale.
    }
    \label{fig:detector}
\end{figure}

As outlined in the previous section, the LDMX detector concept consists of three main parts: a tracking system to measure the transverse momentum, an ECal to determine the energy, and an HCal as a veto instrument. The conceptual design is depicted in Fig.~\ref{fig:detector}. 

The nominal target is a tungsten sheet with a thickness corresponding to 0.1 radiation lengths ($X_0$). Located upstream of the target inside a 1.5\,T magnetic field is the tagging tracker, consisting of seven silicon-strip modules. Downstream of the target, in the fringe field of the magnet, a recoil tracker based on similar silicon-strip modules enables tracking of charged particles with $p\geq50$\,MeV.  
The ECal is a highly-granular Si-W sampling calorimeter with 34 layers and a total depth of $40\,X_0$. This depth enables the full containment and precise measurement of electromagnetic showers. The 
high granularity, with cells of about $0.5\mathrm{\,cm}^2$, makes it possible to distinguish electromagnetic showers from other energy depositions, such as hadronic products of photo-nuclear reactions and tracks from minimum-ionising particles. 
The ECal is surrounded  by the large HCal, except on the beam-facing side. The HCal is a sampling calorimeter with steel absorber layers and scintillator bars as active material, read out via silicon photomultipliers (SiPMs). Its depth corresponds to about $17\lambda_A$ (nuclear interaction lengths) in the forward direction. It is designed to contain highly-penetrating neutral hadrons and capture scattering products at large angles. 
The main physics trigger is the missing energy signature, obtained from a fast measurement of the energy deposited in the first 20 layers of the ECal. 

\subsection{Background Processes and Veto Handles}
\label{sec:bkgLadder}
\begin{figure}
    \centering
    \includegraphics[width=1.0\textwidth]{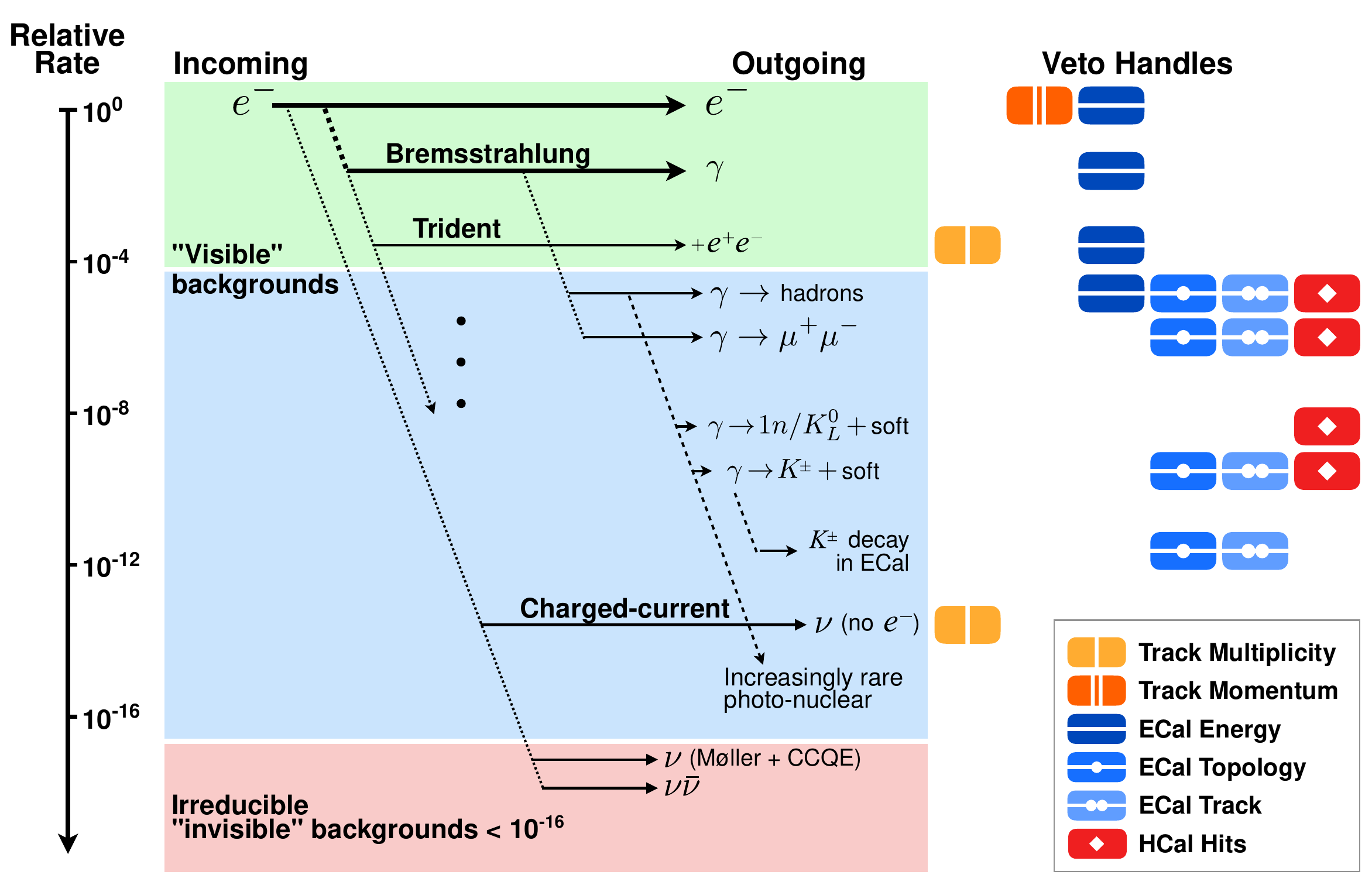}
    \caption{Relative rates of background processes at 8\,GeV beam energy, along with the veto handles used to reject them, down to a relative rate of $< 10^{-16}$. The  colour shading groups different background classes: The upper green section corresponds to events in which the full beam energy is deposited in the ECal; the middle blue section is rarer reactions resulting in anomalous energy depositions in the ECal or particles escaping the ECal; and the bottom red section corresponds to irreducible backgrounds with real missing energy.}
    \label{fig:backgroundLadder}
\end{figure}

As stated in Sec.~\ref{sec:thBkg}, the signature of dark matter production in LDMX consists of missing energy and missing transverse momentum after the electron has interacted in the target. Potential backgrounds to this signature and their rate relative to the rate of incoming electrons are shown in Fig.~\ref{fig:backgroundLadder}. First, it is worth noting that there are no irreducible physics backgrounds, since the rate for neutrino processes generating genuine missing energy lies below the envisaged sensitivity up to O($10^{17}$) EoT. In other words, all relevant backgrounds are instrumental backgrounds, related to
limitations in the efficiencies with which visible reaction products can be detected. For most background processes, LDMX can exploit complementary information from several detector systems, as illustrated on the right of Fig.~\ref{fig:backgroundLadder}. 

The most common class of visible backgrounds are processes in which the full beam energy is deposited in the form of electromagnetic showers in the ECal, including: the electron passing through the target without any significant reaction, the emission of a bremsstrahlung photon, and trident production of an additional $e^+e^-$ pair. 
All of these can be efficiently rejected by a measurement of the deposited energy. In cases of a non-interacting electron or trident production,
tracking information further improves the rejection.

More challenging backgrounds originate from bremsstrahlung events in which the photon does not shower in the ECal, but instead converts into a $\mu^+\mu^-$ or hadron pair, or undergoes other photo-nuclear reactions. These photon-induced processes are the focus of this paper.  
Muon conversion events in the ECal can be rejected by several different veto handles. 
The high-granularity of the ECal enables the observation of tracks that sufficiently energetic muons leave as minimum ionising particles (MIP). 
Moreover, the muons will also deposit energy in the HCal, allowing for an additional veto handle to reject this background. 
In case the muon conversion happens in the target, the additional tracks in the recoil tracker provide a veto handle.  

Several orders of magnitude rarer are photo-nuclear reactions resulting in the production of a single energetic neutral hadron that is accompanied by only low-energy particles. Vetoing these backgrounds relies mainly on the HCal, designed to achieve an inefficiency of less than $10^{-6}$ for detecting few-GeV neutrons. 
The production of single charged kaons again occurs at a slightly lower rate and can be rejected using event topology and MIP track information from the ECal, as well as activity in the HCal. An exception are cases in which the kaon decays inside the ECal, and its visible decay products, i.e. particles other than neutrinos, are contained in the ECal. In this case, the ECal information is the only handle to reject these processes. The rate of single charged kaons decaying inside the ECal is expected to be reduced by a factor of around two, compared to 4\,GeV beam energy, due to the larger distance typically traveled by the more energetic kaons.

\section{Simulated Event Samples}
\label{sec:sim}
The simulations employed in this paper are based on the same software suite as described in Ref.~\cite{VetoPaper}, 
 using a customised version of the \geant toolkit (\texttt{10.02.p03}).
The simulation of signal events is done in two steps: First, the interaction of the 8\,GeV electron with a tungsten nucleus is generated with MadGraph/MadEvent~\cite{alwall:2007st}, using a dark photon model originally used for the APEX test run~\cite{Abrahamyan:2011gv} and the HPS experiment~\cite{Adrian:2018scb}, which has been extended to include light DM particles. 
For this analysis, simulated event samples for dark photon masses of 1\,MeV, 10\,MeV, 100\,MeV, and 1\,GeV are generated. 
The resulting particles, i.e. mainly the deflected electron, are then passed to \geant to model the detector response.
These events are used to both develop the analysis selection and to evaluate the signal efficiency.
Each event is simulated with exactly one incoming beam electron, such that no pile-up is present. 

The simulation of background events is done purely in \geant, including the same modifications as described in Ref.~\cite{VetoPaper} to better model some of the design-driving final states. As motivated in Sec.~\ref{sec:bkgLadder}, the four following background classes are considered: photo-nuclear reactions and photon conversion to a $\mu^+\mu^-$ pair, each of which may occur either in the target or in the ECal itself.  
Using dedicated event filters, separate simulated data sets are prepared for these four processes. All require the electron to emit a hard bremsstrahlung-photon of at least 5\,GeV in the target. They are then further filtered according to the fate of the photon by selecting either photo-nuclear reactions or the conversion into two muons from the \texttt{FTFP\_BERT} physics list of \geant, augmented with the \texttt{G4GammaConversionToMuons} list.

The four background samples are listed in Tab.~\ref{tab:bkg_samples} together with the corresponding number of events generated and the equivalent number of EoT. The latter differs from the number of simulated events due to the use of the \geant occurrence biasing method. This allows the generation of high-statistics samples of rare processes more efficiently, by modifying the interaction probabilities such that the cross-section for a specific process can be enhanced. For the simulations in this paper, the photo-nuclear interaction cross-section was increased by a factor of 550 and the muon-conversion cross-section by a factor of 10000 for the conversion happening in the target, and 30000 for the conversion in the ECal.  
A large sample of unbiased events is produced to validate that the position distribution of where reactions occur in the detector in biased events still have a profile consistent with the original simulation.
Each event is then assigned a weight to account for this change and ensure that the total cross section remains constant. 
These biasing techniques allowed for the simulation of an ECal PN sample equivalent to $1.98\times10^{14}$ EoT, as listed in Tab.~\ref{tab:bkg_samples}, and further efforts are required to reach a $10^{16}$ EoT sample, as expected for the full 8\,GeV run. 

\begin{table}[tbp]
    \centering
    \begin{tabular}{  r | c | c  } 
        \hline \hline
        \textbf{Simulated sample} & \textbf{Total events simulated} & \textbf{EoT equivalent} \\
        \hline \hline 
        \ecal \pn                      & $3.60\times 10^{11}$ & $1.98 \times 10^{14}$ \\
        \ecal $\gamma \to \mu \mu$     & $8.00\times10^{10}$ & $2.40\times10^{15}$ \\
        Target \pn                     & $1.63\times10^{12}$ & $8.99\times10^{14}$ \\
        Target $\gamma \to \mu \mu$    & $9.45\times10^{11}$ & $9.45\times10^{15}$ \\
        \hline
    \end{tabular}
    \caption{\label{tab:bkg_samples}
    The background simulations used in this analysis along with the corresponding statistics.  The  detector volume (ECal or Target) specifies where the interaction is simulated. 
    The number of EoT is estimated from the number of simulated events, taking the biasing factor into account.  
    } 
\end{table}

For signal events and the main background, PN reactions occurring in the ECal, additional smaller samples with 4\,GeV beam energy were generated for comparison with the larger 8\,GeV data set.
All background simulations except for the ECal PN background at 8\,GeV were performed using the Lightweight Distributed Computing System developed for LDMX~\cite{LDCS}.

\section{Comparisons Between Different Beam Energies}
\label{sec:energycomparison}
This section showcases some of the differences resulting from an increase in the beam energy. We focus on the main background, PN processes occurring in the ECal, and the signal for a dark-photon mass of 1\,MeV. Higher masses will be shown to generally exhibit larger differences to the background. All of the distributions shown include only events that are retained by the missing energy trigger, which places a requirement on the maximum amount of energy deposited in the 20 front layers of the ECal. For details on the trigger see Sec.~\ref{subsec:trig}. Comparisons for additional variables can be found in App.~\ref{app:energyComparison}.

\begin{figure}
    \centering    
    \smaller
    \def\svgwidth{0.9\textwidth}
    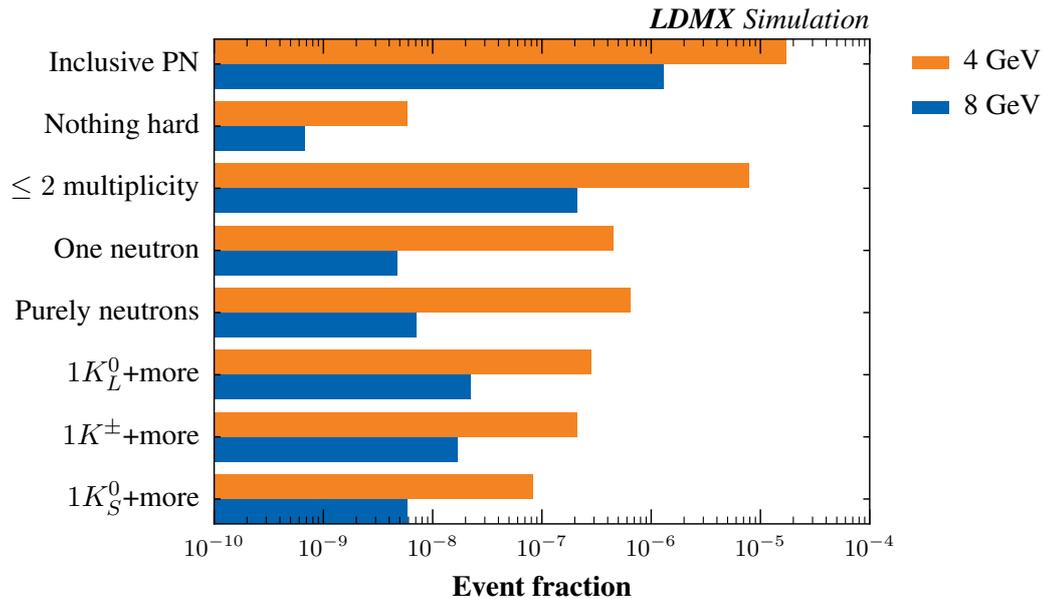
    \caption{Classification of daughter particles from bremsstrahlung photon PN interactions in the ECal based only on particles with at least 200 MeV kinetic energy. Events passing the trigger are shown as a fraction of EoT. The kaon rows consider exactly one kaon and any number of other particles, and the categories are not required to be exclusive.} 
    \label{fig:PNType}
\end{figure}

The most challenging ECal PN reactions to identify are those resulting in low-multiplicity final states of either long-lived neutral particles or charged particles that decay with most of their energy being carried away by neutrinos. 
Fig. 4 illustrates the reduction in the relative number of PN events expected when moving from a 4 to 8 GeV beam energy. The relative frequencies of specific final states are shown, based on particles with 200 MeV or more kinetic energy.
 The number of such photo-nuclear events passing the trigger requirement generally decreases by roughly an order of magnitude. 
 A similar decrease is seen in the {\it nothing hard} category, where there are no particles above 200 MeV in the final state. A larger relative decrease is seen in the low-multiplicity bin, with two or fewer particles above 200 MeV in the final state, and the pure neutron categories that rely on the HCal veto performance. 

The rate of $K^{0}_{L}$ events, which in their most challenging form may decay inside the ECal, with most of the energy being carried away by a neutrino, is similarly more rare at 8 GeV. The apparent higher rate for $K^{0}_{L}$ production compared to $K^{0}_{S}$ is a trigger effect, explained by $K^{0}_{L}$ events depositing energy deeper into the ECal in layers not considered by the trigger, unlike the promptly decaying $K^{0}_{S}$. 

\begin{figure}
    \centering
    \smaller
    \def\svgwidth{0.97\columnwidth}
    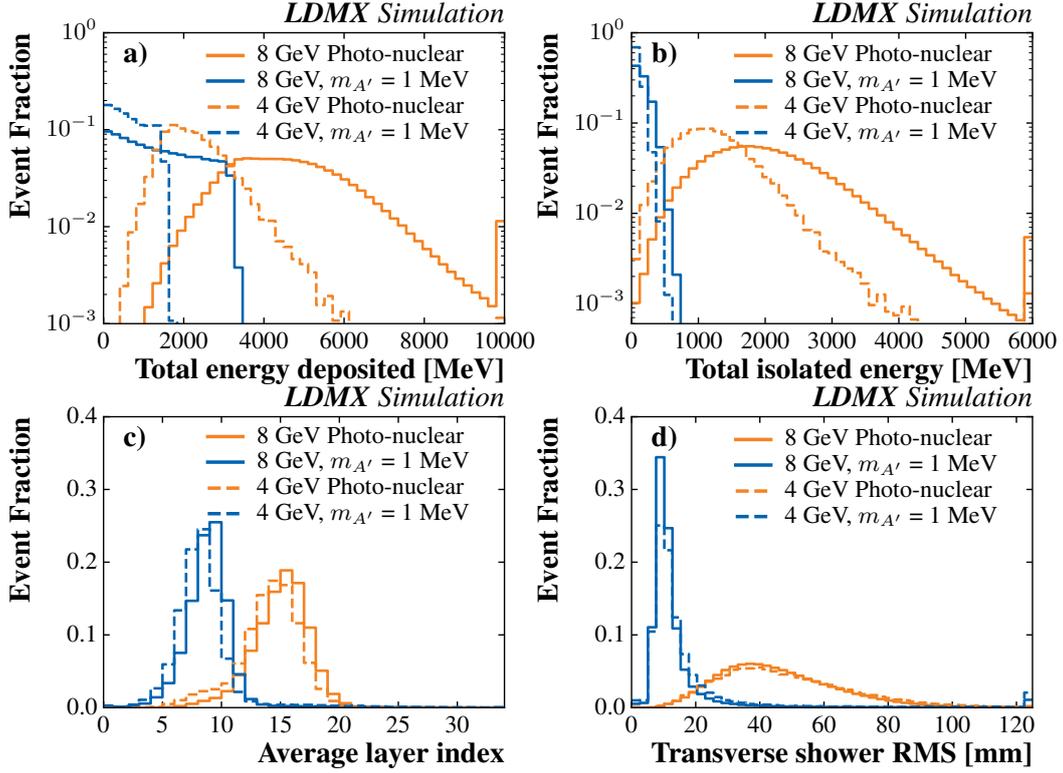
    \caption{Distributions of key variables for dark photon signal events (blue) and ECal PN background events (orange) at 8\,GeV (solid) and 4\,GeV (dashed). All distributions are normalised to the number of events in the respective sample after the trigger requirement. a)  The summed energy of all ECal hits. Due to the experimental resolution, the total energy can sometimes exceed the energy of the incoming beam electron. b) The energy in isolated hits, with no energy deposits in neighboring ECal cells. c) The average layer index of hits.
    d) The transverse spread of energy depositions in the ECal. 
    The rightmost bins are overflow bins. 
    \vspace{0.1in}
    } \label{fig:quadEnergyComparison}
\end{figure}

Fig.~\ref{fig:quadEnergyComparison} shows a comparison of several variables from activity in the ECal, in events passing the trigger requirement, that can be used to discriminate between a dark matter signal and the main PN background, for different beam energies. 
The total amount of reconstructed energy deposited in the ECal is typically higher at 8\,GeV, as is to be expected. For both beam energies, the signal is characterised by lower energy depositions than the background. The signal distributions are truncated by the trigger selection, since most of the energy is deposited close to the front of the ECal in signal events. For the PN background, a considerable fraction of the energy might be deposited deeper into the ECal, such that the total amount of energy can be greater than the energy threshold applied at the trigger level. 

The total isolated energy is defined as the energy deposited in calorimeter cells that have no neighboring cell with significant energy depositions around them in the same silicon layer. This variable helps to capture features separated from the electron shower, in particular diffuse hadronic showers that might originate from PN reaction products. 
It is observed that the amount of energy in such isolated cells increases with higher beam energy. The background is characterised by more energy being deposited in isolated cells, i.e. the energy deposits from the background process are indeed more sparse compared to the dense showers from the recoil electron in signal events. 
At 4\,GeV beam energy, this variable was found to be one of the most discriminating ones~\cite{VetoPaper}, and the distributions in Fig.~\ref{fig:quadEnergyComparison} suggest a similar discrimination power at 8\,GeV.

Fig.~\ref{fig:quadEnergyComparison} further shows the distribution of the average ECal layer of hits. For each event, an average is computed by weighting the layer number with the amount of energy deposited in it. This variable is sensitive to differences in the longitudinal profile between signal and background events, exploiting the fact that the hadronic activity from PN reactions will penetrate deeper into the ECal than the recoil electron shower. This is reflected in a higher average layer number for PN events compared to signal at both beam energies. There is a slight shift to higher values when going from 4\,GeV to 8\,GeV, indicating that showers penetrate deeper into the ECal, as expected for more energetic particles. 

Lastly, the RMS of the transverse positions of ECal hits exploits differences in the lateral profile.
It is significantly larger for the background than for the signal, due to the additional, often hadronic, activity from the PN reactions further away from the electron shower. It is observed that the transverse profile of the electron shower in signal events become slightly narrower at 8\,GeV. The background has a similar distribution at both beam energies, as this variable is dominated by the separation between the electron shower and the PN reaction products, rather than the individual shower profiles.

\section{Veto Methodology and Performance}
\label{sec:evtSel}
In this section, the veto handles outlined in Sec.~\ref{sec:bkgLadder} are described in more detail and their performance is quantified based on the simulations summarised in Sec.~\ref{sec:sim}. The analysis consists of missing energy and missing momentum conditions, a veto based on the topology of energy deposits in the ECal, an independent HCal activity veto, and a fine-grained MIP tracking step to address certain rare photo-nuclear backgrounds. 

\subsection{Trigger and Kinematic Selection}
\label{subsec:trig}
The vast majority of ECal showers started by the signal electron will be contained within the front of the ECal, 
and whether there is significant missing energy will be evident even if the whole ECal volume is not considered. 
In this study, the first 20 layers of the ECal are considered in the calculation of the deposited energy for the trigger decision. Only events with less than 3160\,MeV deposited in this detector region are retained for further analysis, corresponding to a missing energy requirement of 4840\,MeV.
This value was chosen to match the trigger efficiency for the signal obtained in the 4\,GeV analysis. There, 1500\,MeV of deposited energy in the front ECal was required instead, a change roughly consistent with the factor two difference in beam energy. 

Following the trigger selection, a more stringent missing energy criterion with all ECal layers included is introduced, 
with the complete information from the ECal that would be available offline. Using the same numerical value as in the previous online trigger decision, events with more than 3160\,MeV total energy in the whole ECal will be vetoed.

Additionally, events are required to have exactly one track in the recoil tracker with a momentum less than 2400\,MeV/c, 
to ensure that momentum was lost in the target.
This is twice the value used at 4\,GeV beam energy.

\subsection{ECal Shower Shape Veto}

\begin{figure}
    \centering    
    \smaller
    \def\svgwidth{1.0\textwidth}
    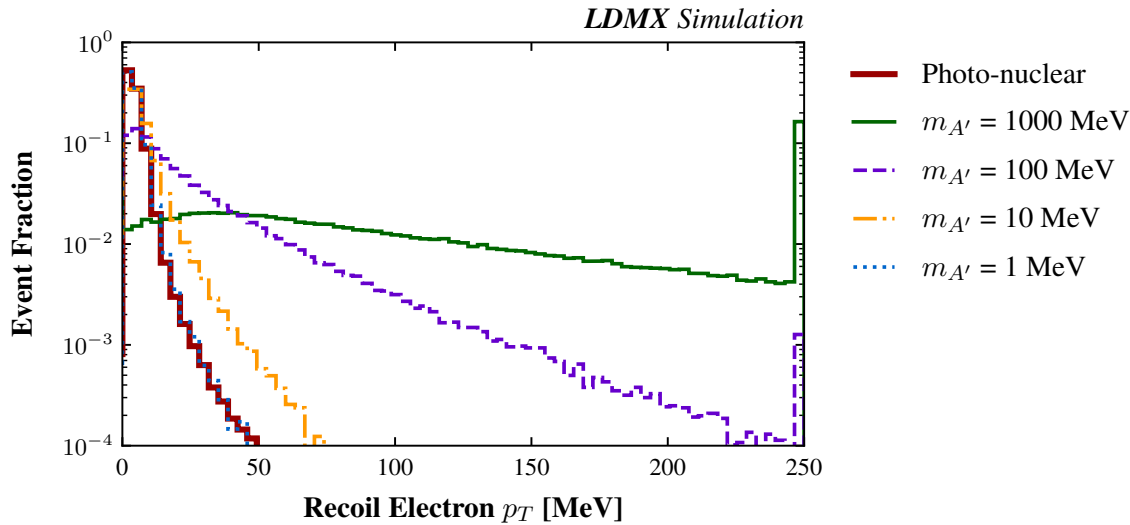
    \caption{$p_{T}$ distribution of the recoiling electron in an 8\,GeV beam, in events passing the missing energy trigger, for dark photon signal events at various mediator masses $m_{A'}$, and background events with photo-nuclear reactions in the ECal.  The rightmost bin is an overflow bin.
    } 
    \label{fig:pTSpectra}
\end{figure}
\begin{figure}

    \hspace{4.1in}{\textbf{\textit{LDMX}} \textit{Simulation}}
    \centering
    
    \includegraphics[width=0.85\textwidth]{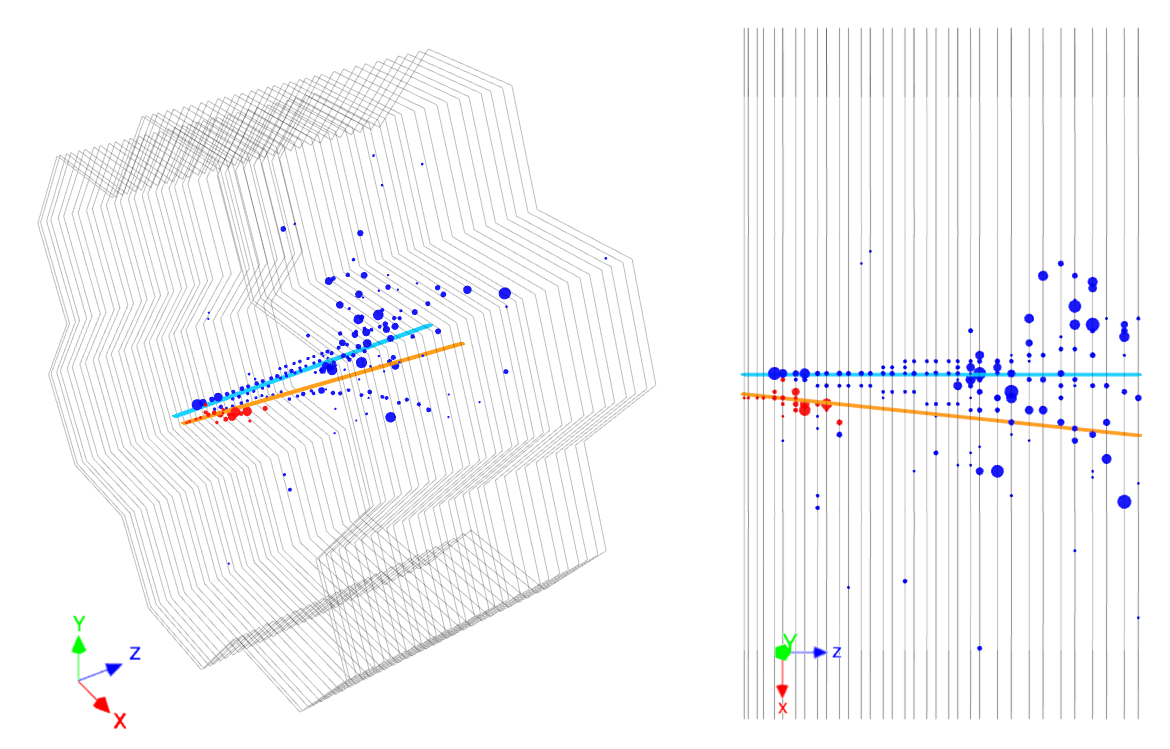}
    \caption{Event display of ECal activity for a photo-nuclear event vetoed by the BDT. A small electron shower (red) is deposited in the first few ECal layers along the expected trajectory of the recoil electron (orange). Proton and charged pion showers (blue) induced by the bremsstrahlung photon are deposited deeper in the ECal, with significant deposits seen near the expected photon trajectory (cyan).
    }
    \label{fig:typicalEvent}
\end{figure}

Background events selected by the trigger and not rejected by the track-momentum requirement will predominantly be events in which the electron loses most of its energy to a bremsstrahlung photon, which does not deposit all of its energy visibly in the ECal. One critical channel for photon energy loss is PN reactions, as discussed in Sec.~\ref{sec:bkgLadder}. 
The granularity of the ECal is utilized to discriminate whether the recoil electron shower, which is the only feature present in signal events, is accompanied by other, photon-induced energy depositions.  
The $p_{T}$ distribution of the recoiling electron for ECal photo-nuclear events is shown in Fig.~\ref{fig:pTSpectra}, along with dark photon signal events with different dark photon mediator masses. The bremsstrahlung background and signal events with lighter mediator masses $m_{A'}$ both have similar low $p_{T}$ distributions, with less spatial separation between the electron shower and recoiling photon in the forward ECal than in more massive $m_{A'}$ signal events. 

As exemplified in Fig.~\ref{fig:quadEnergyComparison}, there are many variables providing discrimination between signal and PN backgrounds. To comprehensively exploit these differences, a boosted decision tree (BDT) classifier is used. It is trained on $10^6$ signal events, containing equal amounts of the four simulated dark photon masses, as well as a set of $10^6$ ECal PN events. 
Fig.~\ref{fig:typicalEvent} shows a typical PN event that the BDT classifies as background. It displays pion and a proton showers resulting from the PN reaction, clearly discernible from the short recoil electron shower.

In the following, we first introduce the variables used as input to the BDT and show some of their distributions. Then, we present the performance of the BDT in terms of background rejection and signal efficiency.

\subsubsection{Input Variables}
\begin{figure}
    \centering    
    \smaller
    \def\svgwidth{5in}
    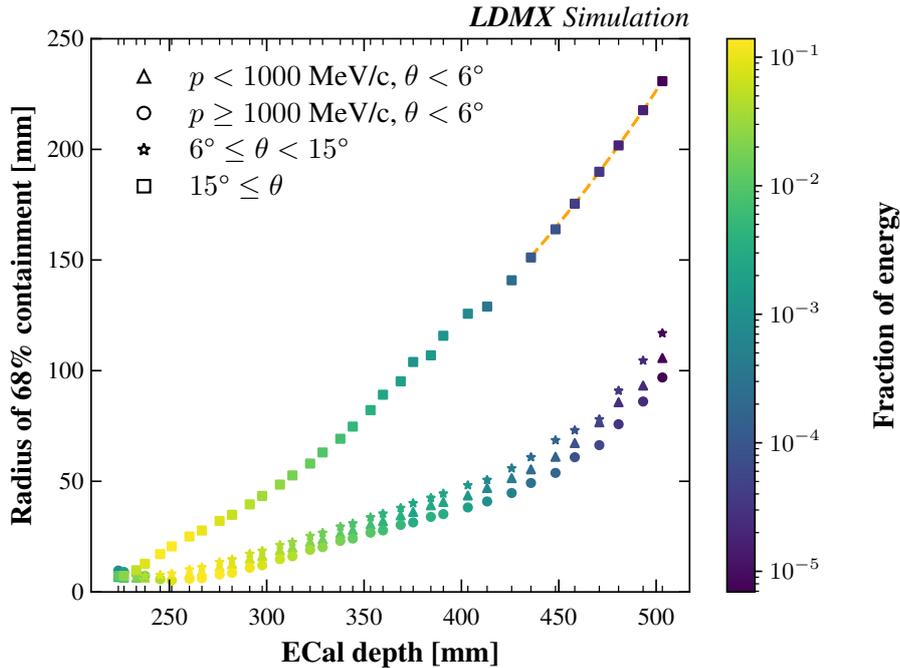
    \caption{The containment radius in each ECal layer is defined as the radius containing on average 68\% of a signal electron's electromagnetic shower. The phase space of electron momentum $p$ and electron angle to the beamline at the target $\theta$ is split into four regions, corresponding to the four curves. In the last five layers for the high angle electrons, a polynomial fit is used to extrapolate the containment radii due to low statistics. The relative distribution of energy between layers, in a certain category of radii, is expressed by the color of the scattered points.
    } 
    \label{fig:containmentradii}
\end{figure}
The purpose of the BDT is to leverage topological features of the energy depositions in the ECal to discriminate signal from the challenging ECal PN backgrounds. For the sake of comparability, the same 42 input variables as in Ref.~\cite{VetoPaper} are used, with some adjustments for the increased beam energy.  They fall into three categories: global features, such as the sum of energy in isolated cells, features describing the transverse or longitudinal distribution of energy in the whole ECal volume, and features describing the energy distribution around the projected electron and photon trajectories. 
The latter is inferred from the initial momentum of the electron before the target, and the trajectory of the electron measured in the recoil tracker. 

For each ECal layer, a radius is defined such that on average 68\% of the energy of an electromagnetic shower in this layer is contained within this radius. These \emph{containment radii} depend on the incident angle and momentum of the recoil electron at the ECal face, and four sets of radii are computed to coarsely capture the change in shower shape development depending on the recoil electron kinematics, resembling the binning used in the 4\,GeV analysis. The containment radii are shown in Fig.~\ref{fig:containmentradii}.  For angles less than $15\degree$, the radii are similar in a given layer, while electrons leaving the target with large angles with respect to the beam direction need considerably larger containment radii.
For electrons, the four sets of containment radii shown in Fig.~\ref{fig:containmentradii} are used.
Around the photon trajectory, the radii for $p \leq 1000$\,MeV and $\theta < 6\degree$ are used.

A set of input variables to the BDT is defined based on the distribution of energy and hits inside and outside the containment radii around the electron and bremsstrahlung photon path.  
These variables include the energy within the electron and photon containment radii, as well as the number of hits and the transverse spread of hits that are well separated from both the electron and the photon paths. 
In addition, the energy and hit distribution in volumes determined by multiples of the containment radius, from one to five containment radii, is used to give a more granular description of the transverse energy distribution.

\begin{figure}
    \centering
    \smaller
    \def\svgwidth{0.95\columnwidth}
    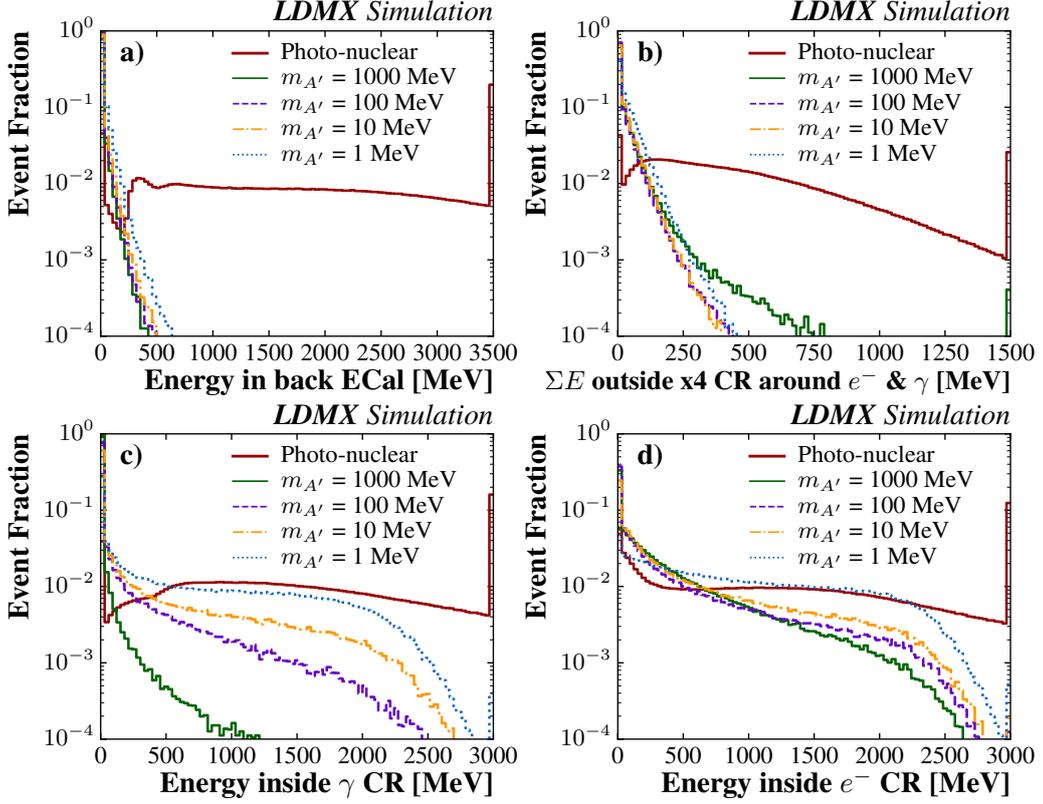
    
    \caption{A selection of BDT input variables. a) Energy in the last 14 ECal layers. b) The total energy (in all ECal layers) outside four times the containment radii around the electron and bremsstrahlung photon trajectories. c) The energy within the containment radii around the photon bremsstrahlung photon trajectory, summed over all ECal layers.
    d) The energy within the containment radii around the electron trajectory, summed over all ECal layers. 
    The rightmost bins are overflow bins.}
    \label{fig:quad2}
\end{figure}

Fig.~\ref{fig:quad2} shows the distributions of some of the BDT input variables for the PN background and four different signal masses; additional distributions can be found in App.~\ref{app:BDTVariables}. The energy deposited in the back 14 layers of the ECal that are not considered for the trigger, and the energy sum of hits outside a region corresponding to four containment radii around both the photon and the electron paths, are both variables that provide good separation of signal and background. The separation is less pronounced when comparing the energy inside the photon and electron containment radii.

\subsubsection{BDT Performance}
\begin{figure}
    \centering    
    \smaller
    \def\svgwidth{4.5in}
    \input{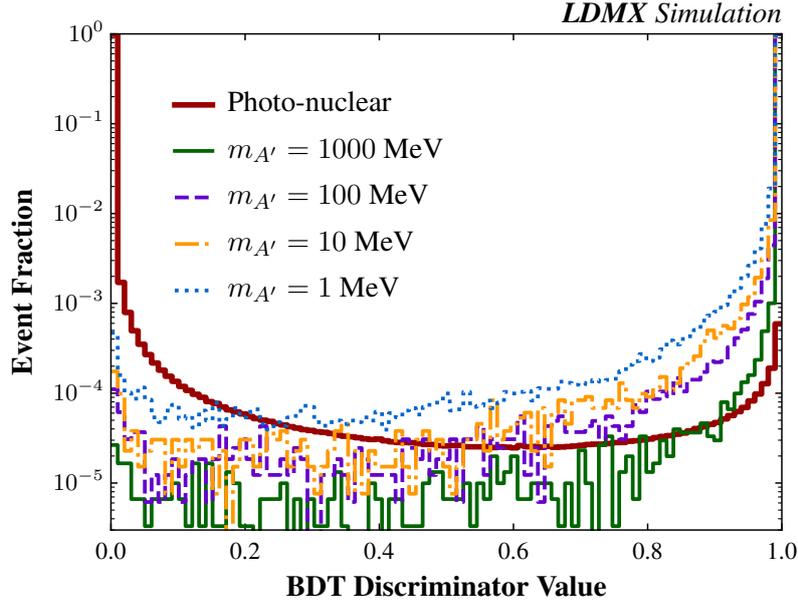}
    \caption{BDT scores assigned to ECal photo-nuclear background events and signal events for various mediator masses.} 
    \label{fig:scoreSeparation}
\end{figure}

\begin{figure}[ht!]
    \centering    
    \smaller
    \def\svgwidth{6.4in}
    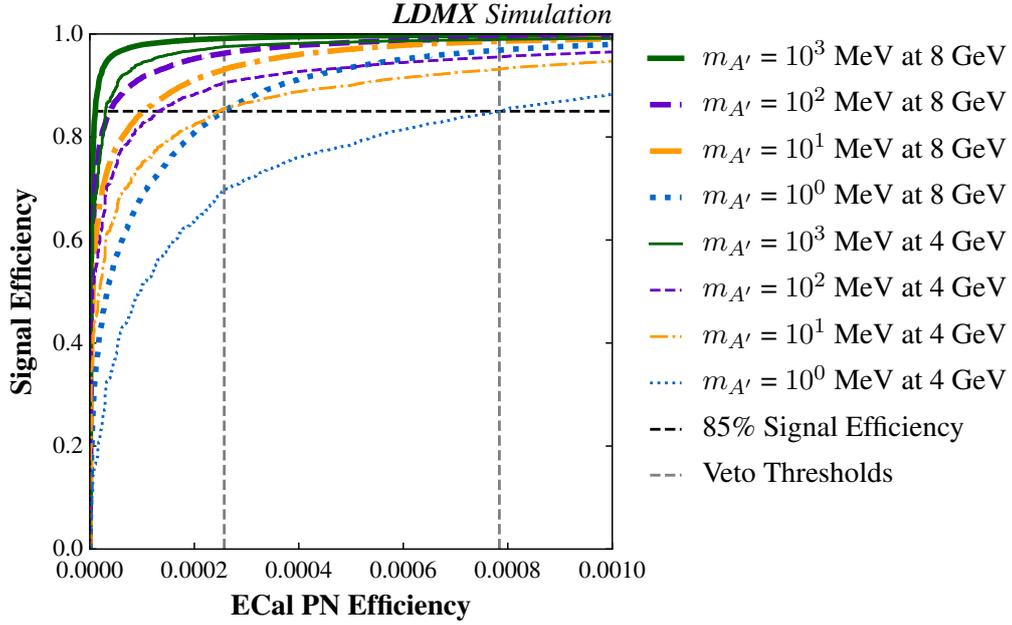
    \caption{ 
    Relationship between background efficiency (of the ECal PN background) and the signal efficiency for different dark photon mediator masses $m_{A'}$. Thicker lines show the performance at 8\,GeV, thinner lines at 4\,GeV. The two vertical dashed grey lines show the background rejection power at 8 GeV and 4\,GeV, from left to right respectively, that maintains at least 85\% signal efficiency across the $m_{A'}$ mass range.} 
    \label{fig:ROC}
\end{figure}

The BDT achieves good separation between signal and background for all considered mediator masses, as displayed in Fig.~\ref{fig:scoreSeparation}. The discriminator value is a score between 0 and 1, where 0 is defined as background-like, and 1 as signal-like.
Fig.~\ref{fig:ROC} shows the relationship between signal and PN background efficiency when varying the requirement on the BDT score.
The similarly trained BDT used in the 4\,GeV analysis is shown alongside the results at 8\,GeV for comparison. 
The same BDT trained on and applied to 8\,GeV data obtains better performance than was true for 4\,GeV.
A requirement on the BDT discriminator value is used to reject PN background events, and the threshold value is chosen to maintain 85\% signal efficiency for the 1\,MeV mediator mass, similar to the efficiency used at 4\,GeV. The 1\,MeV mass point is chosen as the benchmark since it is the most difficult to distinguish from background.
The corresponding background efficiency is $2.6\times10^{-4}$,
 illustrated in Fig.~\ref{fig:ROC}. 
 The fraction of ECal PN events remaining after this veto at 8\,GeV, while requiring the same 1\,MeV signal efficiency, is approximately a factor of three smaller than at 4\,GeV.

\subsection{HCal Activity Veto}
\begin{figure}
    \centering
    \smaller
    \def\svgwidth{0.75\columnwidth}
    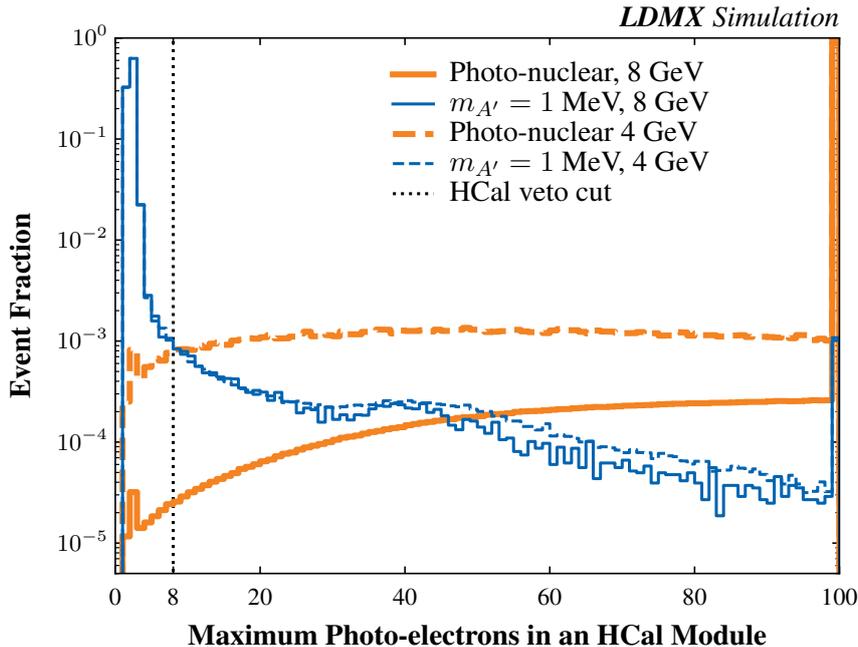
    \caption{The maximum number of photo-electrons measured in any HCal bar, in events passing only the missing energy trigger veto, at both beam energies of 4\,GeV (dashed lines) and 8\,GeV (solid lines). Photo-nuclear backgrounds are shown in orange, and the signal for a 1\,MeV dark photon in blue.  The rightmost bin is an overflow bin. }
    \label{fig:peSeparationBoth}
\end{figure}

Some background events leave little activity in the ECal apart from the recoil electron shower, such as events in which only a single neutron is produced in the PN reaction. These events are very similar to signal events in the ECal, and will typically not be vetoed by the ECal BDT score. They will, however, in most cases produce significant activity in the HCal. 
In contrast, most of the activity will be contained in the ECal for signal events. Energy depositions in the HCal can thus be used as an additional handle to reject background events. 
The HCal veto is defined to reject an event when any scintillator bar in the HCal registers a signal corresponding to 8 or more photo-electrons (PEs) in one SiPM for side HCal bars with single-ended readout, or 8 PEs in total from the two SiPMs at opposite read-out ends for the back HCal bars. This requirement lies well above the expected noise level of 1\,PE equivalent with a tail at the level of $10^{-6}$ extending to at most 6\,PE equivalents.  Fig.~\ref{fig:peSeparationBoth} shows the maximum number of PEs detected in any HCal bar for the PN background and for $m_{A'} = 1$ MeV signal events, at both 4\,GeV and 8\,GeV. A strong discrimination between signal and background is seen, that is considerably more pronounced at 8\,GeV, with at most a few percent of signal events rejected by this criterion. 

\begin{figure}
    \centering
    \smaller
    \def\svgwidth{1.0\columnwidth}
    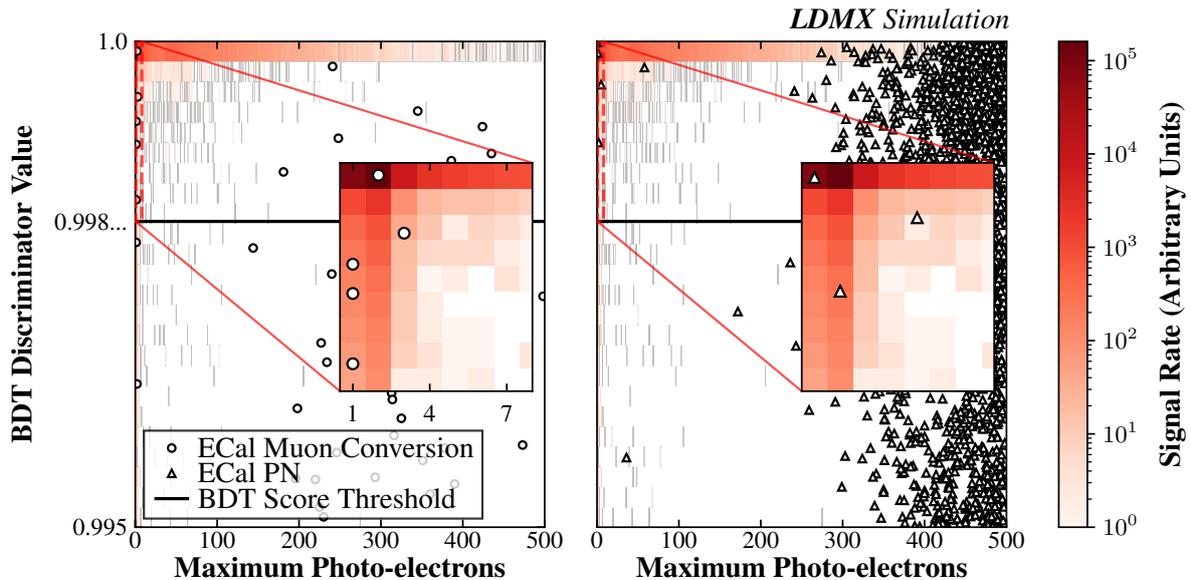
    \caption{Distribution of BDT discriminator values and maximum number of photo-electrons in the HCal, in muon conversion (left) and photo-nuclear (right) events in the ECal, near the signal region. 
    The insets show the signal region, where a few events in both background classes pass both the BDT and HCal vetos.
    The signal rate for $m_{A'} = 1000$ MeV is shown underneath.
    }
    \label{fig:muonconversionPE}
\end{figure}

Fig.~\ref{fig:muonconversionPE} shows the BDT discriminator value versus the maximum number of PEs in any HCal bar for a $m_{A'} = 1$\,GeV signal, compared to $\gamma\rightarrow\mu\mu$ conversion events and PN processes in the ECal. The signal, as expected, accumulates at high BDT values and low HCal activity.  
The insets show a zoomed-in view of the signal region, i.e. events with a BDT score higher than the chosen threshold and at most 8 PEs in any HCal bar. Only three simulated PN events and five simulated muon conversion events are found in the signal region. 
A plot showing the distribution of ECal PN events further away from the signal region can be found in App.~\ref{app:2dplot}.

\begin{figure}[ht]
    \centering
    \smaller
    \def\svgwidth{0.93\columnwidth}
    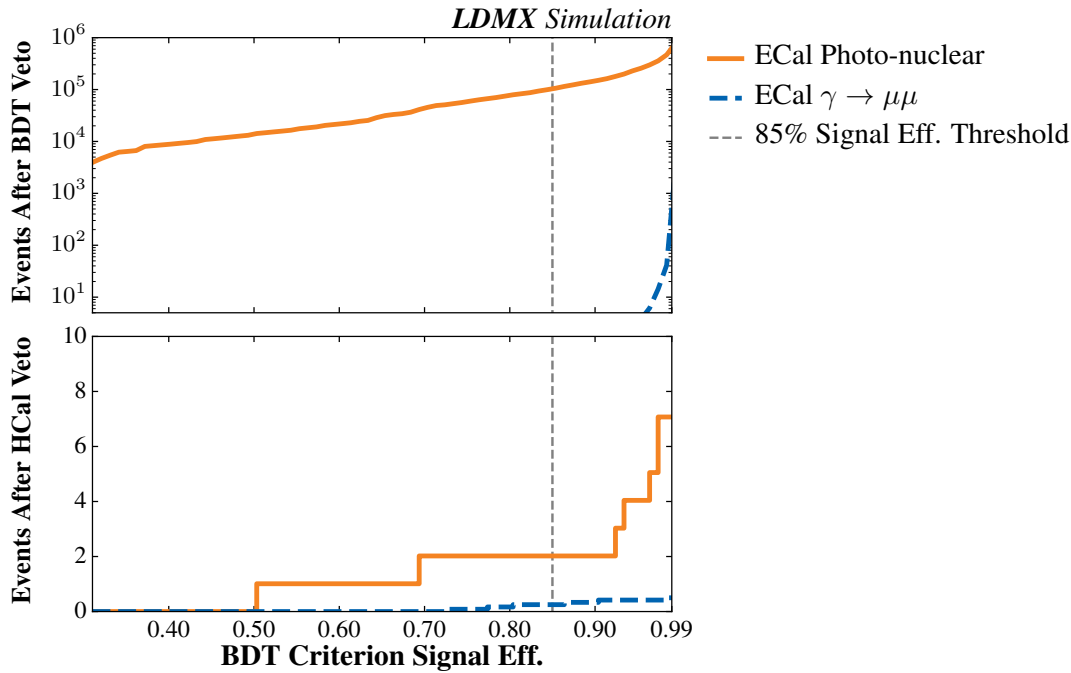
    \caption{The number of events passing up to and including the BDT discriminator value threshold (above) and additionally the HCal activity requirement (below), 
    depending on the chosen BDT signal efficiency for $m_{A'} = 1$ MeV. 
    Both the ECal PN sample (orange) and the  ECal $\gamma\rightarrow \mu\mu$ sample (blue) reach zero events remaining with sufficiently loose signal efficiency requirements. Both backgrounds are scaled to $2\times10^{14}$ EoT in this figure. The performance with an 85\% signal efficiency requirement is marked by the intersections with the dashed grey line.}
    \label{fig:cutvariation}
\end{figure}

Fig.~\ref{fig:cutvariation} further illustrates the BDT and HCal veto complementarity, showing the number of events passing the BDT and subsequent HCal veto, while varying the BDT's required $m_{A'} = 1$ MeV signal efficiency. The ECal $\gamma\rightarrow\mu\mu$ background is largely vetoed by the BDT with large signal efficiencies, unlike the PN background, showing that the ECal is a strong veto handle for the muon conversion background, despite training it on PN events only. While both the $\gamma\rightarrow\mu\mu$ and PN backgrounds have a small number of events passing the additional HCal veto, 
these straggling events are removed at lower required signal efficiencies, meaning that they have potential to be vetoed by energy features in the ECal.

\subsection{MIP Tracking}
\label{subsection:MIPTracking}
Background events that pass both the ECal BDT and the HCal veto will typically have very small, if any, energy depositions in both detector systems. Such signatures can arise from muon conversion if one muon is very soft and the other decays before reaching the HCal, imparting most of its energy to a neutrino and leaving only a short track in the ECal. Similarly, PN reactions can result in the production of charged kaons that decay inside the ECal, leaving only a short track as a detectable signature. 
A dedicated MIP tracking algorithm is used to reject these backgrounds and other topologies with similar challenges. 

First, we search for {\it axial} tracks perpendicular to the ECal face, along the beam-axis in a single row of ECal cells.
Possible hits within one containment radius around the projected recoil electron trajectory are not considered, as activity within that volume is likely to come from the electron shower. If the electron and photon trajectories are sufficiently collinear with a separation of less than 14 degrees, then hits within the containment radius are also considered. The collinear case occurs more frequently at higher beam energies, due to the larger boost. 
If the electron and photon trajectories are very collinear, with a separation less than 8 mm at the ECal face and an angular separation of less than 6 degrees, the two showers will be almost fully overlapping. In such limiting cases, the algorithm searches for short axial tracks outside the containment radius of the electron trajectory, to find features of photo-nuclear products that leave the shower volume, and additionally, it then searches for long axial tracks. The long axial tracks can possibly be near the electron trajectory, to find tracks from very boosted particles that extended further into the detector than the typical electron shower. 
A second step searches for line-like tracks using linear regression, among the ECal hits that are not part of an axial track found so far, to identify MIP tracks from secondaries with a large angle with respect to the incoming bremsstrahlung photon.

\begin{figure}
    \centering    
    \smaller
    \def\svgwidth{0.9\columnwidth}
    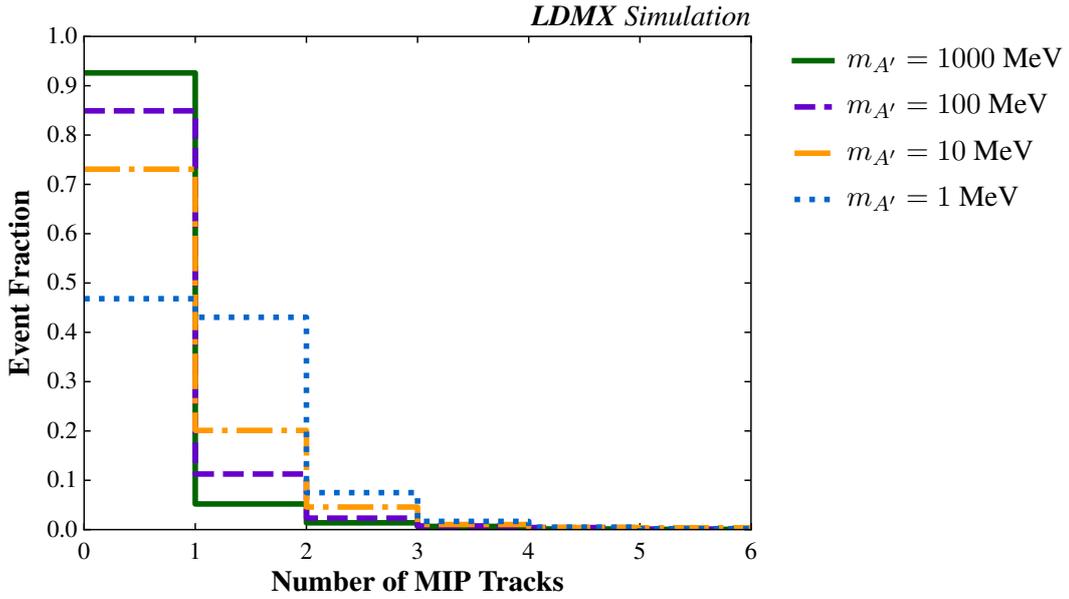
    \caption{The number of MIP tracks found in signal events that pass all previously described vetoes, at various mediator masses.} 
    \label{fig:tracks}
\end{figure}

The number of tracks counted in signal events, that pass all other veto steps, is shown in Fig.~\ref{fig:tracks}. 
Only events with no tracks are considered signal event candidates.
The signal efficiency of this criterion is around 90\% for $m_{A'} = 1$\,GeV, while about 50\% at $m_{A'} = 1$ MeV. 
When the electron and bremsstrahlung trajectories are collinear and produce overlapping showers, it is challenging to resolve MIP tracks, limited by the spatial resolution of the ECal, 
and the algorithm opts for searching near the electron trajectory as well, where fake tracks may be found in hits from the electron shower. 
The electron trajectory and apparent bremsstrahlung trajectory are more collinear in lighter $m_{A'}$ signal events than for heavier mediator masses, and the lower MIP tracking efficiency for $m_{A'} = 1$ MeV is due to fake tracks being found in the recoil electron shower.

\section{Results and Discussion}
\label{sec:res}
The number of background events remaining after applying the rejection criteria described in Sec.~\ref{sec:evtSel} is summarised for all four background samples in Tab.~\ref{tab:cutflowscaled2e14}, scaled to an EoT equivalent of $2\times10^{14}$. A corresponding table with the unscaled event numbers can be found in App.~\ref{app:cutflow}. No background events remain after the full veto sequence.
\begin{table}
\centering
\begin{tabular}{l|c|c|c|c}
\hline \hline
 \multicolumn{1}{l|}{} &  \multicolumn{2}{l|}{{\centering \bf \hfil Photo-nuclear}}    &  \multicolumn{2}{l}{\bf \centering \hfil  Muon conversion}     \\ \cline{2-5}
  &  {\bf \centering Target-area} & {\bf \centering ECal}  & {\bf Target-area} & {\bf ECal}   \\ \hline
 EoT Equivalent & $2.00\times10^{14}$ & $2.00 \times 10^{14}$ & $2.00\times10^{14}$ & $2.00\times10^{14}$  \\ \hline \hline
 Trigger (front ECal energy $ < 3160$ MeV) & $7.57\times10^7$  & $4.43\times10^8$ & $2.37\times10^7$  & $8.12\times10^7$  \\ 
 Total ECal energy $ < 3160$ MeV & $2.73\times10^7$ & $7.27\times10^7$ & $1.76\times10^7$ & $6.06\times10^7$ \\
 Single track with p $< 2400$ MeV/c & $3.03\times10^6$ & $6.64\times10^7$  & $5.32\times10^4$ & $5.69\times10^7$  \\ 
ECal BDT (85\% eff. $m_{A'} = 1$ MeV) & $1.50\times10^5$ & $1.04\times 10^5$ & $< 1$ & $< 1$ \\ 
 HCal max PE $< 8$  & $< 1$ & 2.02  & $< 1$ & $< 1$ \\ 
 ECal MIP tracks = 0 & $< 1$ & $< 1$ & $< 1$ & $< 1$ \\ \hline
\end{tabular}
\caption{Number of simulated background events remaining at the different selection stages, scaled to an EoT equivalent of $2\times10^{14}$. The background classes are separated by where in the detector the reaction took place, either in the target or in the ECal.}
\label{tab:cutflowscaled2e14}
\end{table}

The single-track requirement mainly removes backgrounds originating in the target, as the reaction products can leave additional tracks in the recoil tracker. This requirement is particularly effective for muon-conversion in the target.
For the target PN background, on the other hand, additional charged particles will often leave a sizeable signal in the ECal and could thus be removed by the trigger rather than the subsequent track requirement, or if neutral particles are produced in the target, these will not be seen in the recoil tracker. For the backgrounds in which the photon interacts only in the ECal, the track criterion only removes a small fraction of events in the rare occasions where particles back-scatter from the ECal. 

The BDT reduces the ECal PN background further by more than two orders of magnitude. It is also highly efficient in discarding the muon backgrounds, both from the target and the ECal, leaving less than one event for $2\times10^{14}$ EoT. 
The BDT also reduces the background from PN reactions in the target considerably, yet the HCal veto proves to be most efficient in removing this background, rejecting all events that survive the BDT requirement. The HCal veto also removes all but 2 ECal PN events. This may be compared to results obtained with a beam energy of 4\,GeV, where it was found that 10 events remained for a similar sample size, including a small set of charged kaon events that required the introduction of the MIP tracking. After the HCal veto, all of the muon-conversions events in the target are removed, and all but 0.25 of the ECal muon-conversions (tabulated as $< 1$ in Tab.~\ref{tab:cutflowscaled2e14}). 
The few events surviving after the HCal veto are eventually removed by the MIP tracking algorithm, such that all simulated photon-induced background events are rejected for $2\times10^{14}$\,EoT at a beam energy of 8\,GeV. A highly efficient background rejection is then expected for an initial data taking phase of $4\times10^{14}$\,EoT.

\FloatBarrier
\section{Conclusion}
\label{sec:concl}
The Light Dark Matter eXperiment will use a novel fixed-target, missing-momentum approach to search for sub-GeV dark matter and other new physics in forward electron-scattering with comprehensive, model-independent sensitivity.
The main backgrounds arise from photon-induced reactions in which the energy carried by the photon is transferred to a small number of final state particles that interact only very weakly or not at all with the detector. Particularly challenging examples are photo-nuclear reactions resulting in single neutral or charged particles. 

LDMX is being designed for an initial beam energy of 4\,GeV, and the majority of the data will be collected at 8\,GeV. Employing detailed GEANT4-based simulations, this paper shows that the rejection strategy developed for 4\,GeV is successful in rejecting photon-induced backgrounds for the planned initial data samples of $4\times10^{14}$ electrons on target at 8\,GeV, and no new limiting background event types are found to become relevant at 8\,GeV.

\clearpage

\begin{acknowledgments}
Support for UCSB is made possible by the Joe and Pat Yzurdiaga endowed chair in experimental science. 
Use was made of the UCSB computational facilities administered by the Center for Scientific Computing at the California NanoSystems Institute and Materials Research Laboratory (an NSF MRSEC; DMR-1720256) and purchased through NSF CNS-1725797, and from  
resources provided by the Swedish National Infrastructure for Computing at the Centre for Scientific and Technical Computing at Lund University (LUNARC), as well as LUNARC’s own infrastructure. Contributions from Caltech, CMU, Stanford, TTU, UMN, and UVA are supported by the US Department of Energy under grants DE-SC0011925, DE-SC0010118, DE-SC0022083, DE-SC0015592, DE-SC00012069, and DE-SC0007838, respectively. Support for Lund University is made possible by the Knut and Alice Wallenberg foundation (project grant Light Dark Matter, Dnr. KAW 2019.0080), and by the Crafoord foundation (Dnr 20190875) and the Royal Physiographic Society of Lund.  RP acknowledges support through the L’Or\'{e}al-UNESCO For Women in Science in Sweden Prize with support of the Young Academy of Sweden, and from the Swedish Research Council (Dnr 2019-03436). LB acknowledges support from the Knut and Alice Wallenberg Foundation Postdoctoral Scholarship Program at Stanford (Dnr. KAW 2018.0429). JE, GK, CH, WK, CMS, and NT are supported by the Fermi Research Alliance, LLC under Contract No. DE-AC02-07CH11359 with the U.S. Department of Energy, Office of Science, Office of High Energy Physics and the Fermilab LDRD program.  CB, PB, OM, TN, PS, and NT are supported by Stanford University under Contract No. DE-AC02-76SF00515 with the U.S. Department of Energy, Office of Science, Office of High Energy Physics.
\end{acknowledgments}


\bibliography{bibliography}

\begin{thebibliography}{21}%
\makeatletter
\providecommand \@ifxundefined [1]{%
 \@ifx{#1\undefined}
}%
\providecommand \@ifnum [1]{%
 \ifnum #1\expandafter \@firstoftwo
 \else \expandafter \@secondoftwo
 \fi
}%
\providecommand \@ifx [1]{%
 \ifx #1\expandafter \@firstoftwo
 \else \expandafter \@secondoftwo
 \fi
}%
\providecommand \natexlab [1]{#1}%
\providecommand \enquote  [1]{``#1''}%
\providecommand \bibnamefont  [1]{#1}%
\providecommand \bibfnamefont [1]{#1}%
\providecommand \citenamefont [1]{#1}%
\providecommand \href@noop [0]{\@secondoftwo}%
\providecommand \href [0]{\begingroup \@sanitize@url \@href}%
\providecommand \@href[1]{\@@startlink{#1}\@@href}%
\providecommand \@@href[1]{\endgroup#1\@@endlink}%
\providecommand \@sanitize@url [0]{\catcode `\\12\catcode `\$12\catcode
  `\&12\catcode `\#12\catcode `\^12\catcode `\_12\catcode `\%12\relax}%
\providecommand \@@startlink[1]{}%
\providecommand \@@endlink[0]{}%
\providecommand \url  [0]{\begingroup\@sanitize@url \@url }%
\providecommand \@url [1]{\endgroup\@href {#1}{\urlprefix }}%
\providecommand \urlprefix  [0]{URL }%
\providecommand \Eprint [0]{\href }%
\providecommand \doibase [0]{https://doi.org/}%
\providecommand \selectlanguage [0]{\@gobble}%
\providecommand \bibinfo  [0]{\@secondoftwo}%
\providecommand \bibfield  [0]{\@secondoftwo}%
\providecommand \translation [1]{[#1]}%
\providecommand \BibitemOpen [0]{}%
\providecommand \bibitemStop [0]{}%
\providecommand \bibitemNoStop [0]{.\EOS\space}%
\providecommand \EOS [0]{\spacefactor3000\relax}%
\providecommand \BibitemShut  [1]{\csname bibitem#1\endcsname}%
\let\auto@bib@innerbib\@empty
\bibitem [{\citenamefont {Raubenheimer}(2018)}]{Raubenheimer:2018wwc}%
  \BibitemOpen
  \bibfield  {author} {\bibinfo {author} {\bibfnamefont {T.}~\bibnamefont
  {Raubenheimer}},\ }\bibfield  {title} {\bibinfo {title} {{The LCLS-II-HE, A
  High Energy Upgrade of the LCLS-II}},\ }in\ \href
  {https://doi.org/10.18429/JACoW-FLS2018-MOP1WA02} {\emph {\bibinfo
  {booktitle} {{Proceedings, 60th ICFA Advanced Beam Dynamics Workshop on
  Future Light Sources (FLS2018): Shanghai, China, March 5-9, 2018}}}}\
  (\bibinfo {year} {2018})\ p.\ \bibinfo {pages} {MOP1WA02}\BibitemShut
  {NoStop}%
\bibitem [{\citenamefont {Akesson}\ \emph {et~al.}(2018)\citenamefont {Akesson}
  \emph {et~al.}}]{WhitePaper}%
  \BibitemOpen
  \bibfield  {author} {\bibinfo {author} {\bibfnamefont {T.}~\bibnamefont
  {Akesson}} \emph {et~al.} (\bibinfo {collaboration} {LDMX}),\ }\bibfield
  {title} {\bibinfo {title} {{Light Dark Matter eXperiment (LDMX)}},\
  }\href@noop {} {\  (\bibinfo {year} {2018})},\ \Eprint
  {https://arxiv.org/abs/1808.05219} {arXiv:1808.05219 [hep-ex]} \BibitemShut
  {NoStop}%
\bibitem [{\citenamefont {Akesson}\ \emph {et~al.}(2020)\citenamefont {Akesson}
  \emph {et~al.}}]{VetoPaper}%
  \BibitemOpen
  \bibfield  {author} {\bibinfo {author} {\bibfnamefont {T.}~\bibnamefont
  {Akesson}} \emph {et~al.} (\bibinfo {collaboration} {LDMX}),\ }\bibfield
  {title} {\bibinfo {title} {{A high efficiency photon veto for the Light Dark
  Matter eXperiment}},\ }\href {https://doi.org/10.1007/JHEP04(2020)003}
  {\bibfield  {journal} {\bibinfo  {journal} {JHEP}\ }\textbf {\bibinfo
  {volume} {04}},\ \bibinfo {pages} {003}},\ \Eprint
  {https://arxiv.org/abs/1912.05535} {arXiv:1912.05535 [hep-ex]} \BibitemShut
  {NoStop}%
\bibitem [{\citenamefont {Agostinelli}\ \emph {et~al.}(2003)\citenamefont
  {Agostinelli} \emph {et~al.}}]{Agostinelli:2002hh}%
  \BibitemOpen
  \bibfield  {author} {\bibinfo {author} {\bibfnamefont {S.}~\bibnamefont
  {Agostinelli}} \emph {et~al.} (\bibinfo {collaboration} {GEANT4}),\
  }\bibfield  {title} {\bibinfo {title} {{GEANT4: A Simulation toolkit}},\
  }\href {https://doi.org/10.1016/S0168-9002(03)01368-8} {\bibfield  {journal}
  {\bibinfo  {journal} {Nucl. Instrum. Meth.}\ }\textbf {\bibinfo {volume}
  {A506}},\ \bibinfo {pages} {250} (\bibinfo {year} {2003})}\BibitemShut
  {NoStop}%
\bibitem [{\citenamefont {Allison}\ \emph {et~al.}(2016)\citenamefont {Allison}
  \emph {et~al.}}]{Allison:2016lfl}%
  \BibitemOpen
  \bibfield  {author} {\bibinfo {author} {\bibfnamefont {J.}~\bibnamefont
  {Allison}} \emph {et~al.},\ }\bibfield  {title} {\bibinfo {title} {{Recent
  developments in Geant4}},\ }\href
  {https://doi.org/10.1016/j.nima.2016.06.125} {\bibfield  {journal} {\bibinfo
  {journal} {Nucl. Instrum. Meth.}\ }\textbf {\bibinfo {volume} {A835}},\
  \bibinfo {pages} {186} (\bibinfo {year} {2016})}\BibitemShut {NoStop}%
\bibitem [{\citenamefont {Akesson}\ \emph {et~al.}(2022)\citenamefont {Akesson}
  \emph {et~al.}}]{snowmass22}%
  \BibitemOpen
  \bibfield  {author} {\bibinfo {author} {\bibfnamefont {T.}~\bibnamefont
  {Akesson}} \emph {et~al.} (\bibinfo {collaboration} {LDMX}),\ }\bibfield
  {title} {\bibinfo {title} {{Current Status and Future Prospects for the Light
  Dark Matter eXperiment}},\ }\href@noop {} {\  (\bibinfo {year} {2022})},\
  \Eprint {https://arxiv.org/abs/2203.08192} {arXiv:2203.08192 [hep-ex]}
  \BibitemShut {NoStop}%
\bibitem [{\citenamefont {Berlin}\ \emph {et~al.}(2019)\citenamefont {Berlin},
  \citenamefont {Blinov}, \citenamefont {Krnjaic}, \citenamefont {Schuster},\
  and\ \citenamefont {Toro}}]{Berlin:2018bsc}%
  \BibitemOpen
  \bibfield  {author} {\bibinfo {author} {\bibfnamefont {A.}~\bibnamefont
  {Berlin}}, \bibinfo {author} {\bibfnamefont {N.}~\bibnamefont {Blinov}},
  \bibinfo {author} {\bibfnamefont {G.}~\bibnamefont {Krnjaic}}, \bibinfo
  {author} {\bibfnamefont {P.}~\bibnamefont {Schuster}},\ and\ \bibinfo
  {author} {\bibfnamefont {N.}~\bibnamefont {Toro}},\ }\bibfield  {title}
  {\bibinfo {title} {{Dark Matter, Millicharges, Axion and Scalar Particles,
  Gauge Bosons, and Other New Physics with LDMX}},\ }\href
  {https://doi.org/10.1103/PhysRevD.99.075001} {\bibfield  {journal} {\bibinfo
  {journal} {Phys. Rev. D}\ }\textbf {\bibinfo {volume} {99}},\ \bibinfo
  {pages} {075001} (\bibinfo {year} {2019})},\ \Eprint
  {https://arxiv.org/abs/1807.01730} {arXiv:1807.01730 [hep-ph]} \BibitemShut
  {NoStop}%
\bibitem [{\citenamefont {Izaguirre}\ \emph {et~al.}(2015)\citenamefont
  {Izaguirre}, \citenamefont {Krnjaic}, \citenamefont {Schuster},\ and\
  \citenamefont {Toro}}]{PhysRevLett.115.251301}%
  \BibitemOpen
  \bibfield  {author} {\bibinfo {author} {\bibfnamefont {E.}~\bibnamefont
  {Izaguirre}}, \bibinfo {author} {\bibfnamefont {G.}~\bibnamefont {Krnjaic}},
  \bibinfo {author} {\bibfnamefont {P.}~\bibnamefont {Schuster}},\ and\
  \bibinfo {author} {\bibfnamefont {N.}~\bibnamefont {Toro}},\ }\bibfield
  {title} {\bibinfo {title} {Analyzing the discovery potential for light dark
  matter},\ }\href {https://doi.org/10.1103/PhysRevLett.115.251301} {\bibfield
  {journal} {\bibinfo  {journal} {Phys. Rev. Lett.}\ }\textbf {\bibinfo
  {volume} {115}},\ \bibinfo {pages} {251301} (\bibinfo {year} {2015})},\
  \Eprint {https://arxiv.org/abs/1505.00011} {arXiv:1505.00011} \BibitemShut
  {NoStop}%
\bibitem [{\citenamefont {deNiverville}\ \emph {et~al.}(2017)\citenamefont
  {deNiverville}, \citenamefont {Chen}, \citenamefont {Pospelov},\ and\
  \citenamefont {Ritz}}]{deniverville:2016rqh}%
  \BibitemOpen
  \bibfield  {author} {\bibinfo {author} {\bibfnamefont {P.}~\bibnamefont
  {deNiverville}}, \bibinfo {author} {\bibfnamefont {C.-Y.}\ \bibnamefont
  {Chen}}, \bibinfo {author} {\bibfnamefont {M.}~\bibnamefont {Pospelov}},\
  and\ \bibinfo {author} {\bibfnamefont {A.}~\bibnamefont {Ritz}},\ }\bibfield
  {title} {\bibinfo {title} {{Light dark matter in neutrino beams: production
  modelling and scattering signatures at MiniBooNE, T2K and SHiP}},\ }\href
  {https://doi.org/10.1103/PhysRevD.95.035006} {\bibfield  {journal} {\bibinfo
  {journal} {Phys. Rev. D}\ }\textbf {\bibinfo {volume} {95}},\ \bibinfo
  {pages} {035006} (\bibinfo {year} {2017})},\ \Eprint
  {https://arxiv.org/abs/1609.01770} {arXiv:1609.01770 [hep-ph]} \BibitemShut
  {NoStop}%
\bibitem [{\citenamefont {Lees}\ \emph {et~al.}(2017)\citenamefont {Lees} \emph
  {et~al.}}]{Lees:2017lec}%
  \BibitemOpen
  \bibfield  {author} {\bibinfo {author} {\bibfnamefont {J.~P.}\ \bibnamefont
  {Lees}} \emph {et~al.} (\bibinfo {collaboration} {BaBar}),\ }\bibfield
  {title} {\bibinfo {title} {{Search for Invisible Decays of a Dark Photon
  Produced in ${e}^{+}{e}^{-}$ Collisions at BaBar}},\ }\href
  {https://doi.org/10.1103/PhysRevLett.119.131804} {\bibfield  {journal}
  {\bibinfo  {journal} {Phys. Rev. Lett.}\ }\textbf {\bibinfo {volume} {119}},\
  \bibinfo {pages} {131804} (\bibinfo {year} {2017})},\ \Eprint
  {https://arxiv.org/abs/1702.03327} {arXiv:1702.03327 [hep-ex]} \BibitemShut
  {NoStop}%
\bibitem [{\citenamefont {Aguilar-Arevalo}\ \emph {et~al.}(2017)\citenamefont
  {Aguilar-Arevalo} \emph {et~al.}}]{Aguilar-Arevalo:2017mqx}%
  \BibitemOpen
  \bibfield  {author} {\bibinfo {author} {\bibfnamefont {A.~A.}\ \bibnamefont
  {Aguilar-Arevalo}} \emph {et~al.} (\bibinfo {collaboration} {MiniBooNE}),\
  }\bibfield  {title} {\bibinfo {title} {{Dark Matter Search in a Proton Beam
  Dump with MiniBooNE}},\ }\href
  {https://doi.org/10.1103/PhysRevLett.118.221803} {\bibfield  {journal}
  {\bibinfo  {journal} {Phys. Rev. Lett.}\ }\textbf {\bibinfo {volume} {118}},\
  \bibinfo {pages} {221803} (\bibinfo {year} {2017})},\ \Eprint
  {https://arxiv.org/abs/1702.02688} {arXiv:1702.02688 [hep-ex]} \BibitemShut
  {NoStop}%
\bibitem [{\citenamefont {Banerjee}\ \emph {et~al.}(2019)\citenamefont
  {Banerjee} \emph {et~al.}}]{NA64:2019imj}%
  \BibitemOpen
  \bibfield  {author} {\bibinfo {author} {\bibfnamefont {D.}~\bibnamefont
  {Banerjee}} \emph {et~al.},\ }\bibfield  {title} {\bibinfo {title} {{Dark
  matter search in missing energy events with NA64}},\ }\href
  {https://doi.org/10.1103/PhysRevLett.123.121801} {\bibfield  {journal}
  {\bibinfo  {journal} {Phys. Rev. Lett.}\ }\textbf {\bibinfo {volume} {123}},\
  \bibinfo {pages} {121801} (\bibinfo {year} {2019})},\ \Eprint
  {https://arxiv.org/abs/1906.00176} {arXiv:1906.00176 [hep-ex]} \BibitemShut
  {NoStop}%
\bibitem [{\citenamefont {Batell}\ \emph {et~al.}(2014)\citenamefont {Batell},
  \citenamefont {Essig},\ and\ \citenamefont {Surujon}}]{batell:2014mga}%
  \BibitemOpen
  \bibfield  {author} {\bibinfo {author} {\bibfnamefont {B.}~\bibnamefont
  {Batell}}, \bibinfo {author} {\bibfnamefont {R.}~\bibnamefont {Essig}},\ and\
  \bibinfo {author} {\bibfnamefont {Z.}~\bibnamefont {Surujon}},\ }\bibfield
  {title} {\bibinfo {title} {{Strong Constraints on Sub-GeV Dark Sectors from
  SLAC Beam Dump E137}},\ }\href
  {https://doi.org/10.1103/PhysRevLett.113.171802} {\bibfield  {journal}
  {\bibinfo  {journal} {Phys. Rev. Lett.}\ }\textbf {\bibinfo {volume} {113}},\
  \bibinfo {pages} {171802} (\bibinfo {year} {2014})},\ \Eprint
  {https://arxiv.org/abs/1406.2698} {arXiv:1406.2698 [hep-ph]} \BibitemShut
  {NoStop}%
\bibitem [{\citenamefont {Andreev}\ \emph {et~al.}(2021)\citenamefont {Andreev}
  \emph {et~al.}}]{Andreev:2021fzd}%
  \BibitemOpen
  \bibfield  {author} {\bibinfo {author} {\bibfnamefont {Y.~M.}\ \bibnamefont
  {Andreev}} \emph {et~al.},\ }\bibfield  {title} {\bibinfo {title} {{Improved
  exclusion limit for light dark matter from e+e- annihilation in NA64}},\
  }\href {https://doi.org/10.1103/PhysRevD.104.L091701} {\bibfield  {journal}
  {\bibinfo  {journal} {Phys. Rev. D}\ }\textbf {\bibinfo {volume} {104}},\
  \bibinfo {pages} {L091701} (\bibinfo {year} {2021})},\ \Eprint
  {https://arxiv.org/abs/2108.04195} {arXiv:2108.04195 [hep-ex]} \BibitemShut
  {NoStop}%
\bibitem [{\citenamefont {Akimov}\ \emph {et~al.}(2023)\citenamefont {Akimov}
  \emph {et~al.}}]{COHERENT:2021pvd}%
  \BibitemOpen
  \bibfield  {author} {\bibinfo {author} {\bibfnamefont {D.}~\bibnamefont
  {Akimov}} \emph {et~al.} (\bibinfo {collaboration} {COHERENT}),\ }\bibfield
  {title} {\bibinfo {title} {{First Probe of Sub-GeV Dark Matter beyond the
  Cosmological Expectation with the COHERENT CsI Detector at the SNS}},\ }\href
  {https://doi.org/10.1103/PhysRevLett.130.051803} {\bibfield  {journal}
  {\bibinfo  {journal} {Phys. Rev. Lett.}\ }\textbf {\bibinfo {volume} {130}},\
  \bibinfo {pages} {051803} (\bibinfo {year} {2023})},\ \Eprint
  {https://arxiv.org/abs/2110.11453} {arXiv:2110.11453 [hep-ex]} \BibitemShut
  {NoStop}%
\bibitem [{\citenamefont {Aguilar-Arevalo}\ \emph {et~al.}(2022)\citenamefont
  {Aguilar-Arevalo} \emph {et~al.}}]{CCM:2021leg}%
  \BibitemOpen
  \bibfield  {author} {\bibinfo {author} {\bibfnamefont {A.~A.}\ \bibnamefont
  {Aguilar-Arevalo}} \emph {et~al.} (\bibinfo {collaboration} {CCM}),\
  }\bibfield  {title} {\bibinfo {title} {{First dark matter search results from
  Coherent CAPTAIN-Mills}},\ }\href
  {https://doi.org/10.1103/PhysRevD.106.012001} {\bibfield  {journal} {\bibinfo
   {journal} {Phys. Rev. D}\ }\textbf {\bibinfo {volume} {106}},\ \bibinfo
  {pages} {012001} (\bibinfo {year} {2022})},\ \Eprint
  {https://arxiv.org/abs/2105.14020} {arXiv:2105.14020 [hep-ex]} \BibitemShut
  {NoStop}%
\bibitem [{\citenamefont {Schuster}\ \emph {et~al.}(2022)\citenamefont
  {Schuster}, \citenamefont {Toro},\ and\ \citenamefont
  {Zhou}}]{Schuster:2021mlr}%
  \BibitemOpen
  \bibfield  {author} {\bibinfo {author} {\bibfnamefont {P.}~\bibnamefont
  {Schuster}}, \bibinfo {author} {\bibfnamefont {N.}~\bibnamefont {Toro}},\
  and\ \bibinfo {author} {\bibfnamefont {K.}~\bibnamefont {Zhou}},\ }\bibfield
  {title} {\bibinfo {title} {{Probing invisible vector meson decays with the
  NA64 and LDMX experiments}},\ }\href
  {https://doi.org/10.1103/PhysRevD.105.035036} {\bibfield  {journal} {\bibinfo
   {journal} {Phys. Rev. D}\ }\textbf {\bibinfo {volume} {105}},\ \bibinfo
  {pages} {035036} (\bibinfo {year} {2022})},\ \Eprint
  {https://arxiv.org/abs/2112.02104} {arXiv:2112.02104 [hep-ph]} \BibitemShut
  {NoStop}%
\bibitem [{\citenamefont {Alwall}\ \emph {et~al.}(2007)\citenamefont {Alwall},
  \citenamefont {Demin}, \citenamefont {de~Visscher}, \citenamefont {Frederix},
  \citenamefont {Herquet}, \citenamefont {Maltoni}, \citenamefont {Plehn},
  \citenamefont {Rainwater},\ and\ \citenamefont {Stelzer}}]{alwall:2007st}%
  \BibitemOpen
  \bibfield  {author} {\bibinfo {author} {\bibfnamefont {J.}~\bibnamefont
  {Alwall}}, \bibinfo {author} {\bibfnamefont {P.}~\bibnamefont {Demin}},
  \bibinfo {author} {\bibfnamefont {S.}~\bibnamefont {de~Visscher}}, \bibinfo
  {author} {\bibfnamefont {R.}~\bibnamefont {Frederix}}, \bibinfo {author}
  {\bibfnamefont {M.}~\bibnamefont {Herquet}}, \bibinfo {author} {\bibfnamefont
  {F.}~\bibnamefont {Maltoni}}, \bibinfo {author} {\bibfnamefont
  {T.}~\bibnamefont {Plehn}}, \bibinfo {author} {\bibfnamefont {D.~L.}\
  \bibnamefont {Rainwater}},\ and\ \bibinfo {author} {\bibfnamefont
  {T.}~\bibnamefont {Stelzer}},\ }\bibfield  {title} {\bibinfo {title}
  {{MadGraph/MadEvent v4: The New Web Generation}},\ }\href
  {https://doi.org/10.1088/1126-6708/2007/09/028} {\bibfield  {journal}
  {\bibinfo  {journal} {JHEP}\ }\textbf {\bibinfo {volume} {09}},\ \bibinfo
  {pages} {028}},\ \Eprint {https://arxiv.org/abs/0706.2334} {arXiv:0706.2334
  [hep-ph]} \BibitemShut {NoStop}%
\bibitem [{\citenamefont {Abrahamyan}\ \emph {et~al.}(2011)\citenamefont
  {Abrahamyan} \emph {et~al.}}]{Abrahamyan:2011gv}%
  \BibitemOpen
  \bibfield  {author} {\bibinfo {author} {\bibfnamefont {S.}~\bibnamefont
  {Abrahamyan}} \emph {et~al.} (\bibinfo {collaboration} {APEX}),\ }\bibfield
  {title} {\bibinfo {title} {{Search for a New Gauge Boson in Electron-Nucleus
  Fixed-Target Scattering by the APEX Experiment}},\ }\href
  {https://doi.org/10.1103/PhysRevLett.107.191804} {\bibfield  {journal}
  {\bibinfo  {journal} {Phys. Rev. Lett.}\ }\textbf {\bibinfo {volume} {107}},\
  \bibinfo {pages} {191804} (\bibinfo {year} {2011})},\ \Eprint
  {https://arxiv.org/abs/1108.2750} {arXiv:1108.2750 [hep-ex]} \BibitemShut
  {NoStop}%
\bibitem [{\citenamefont {{\it et al.}}(2018)}]{Adrian:2018scb}%
  \BibitemOpen
  \bibfield  {author} {\bibinfo {author} {\bibfnamefont {P.~H.~A.}\
  \bibnamefont {{\it et al.}}} (\bibinfo {collaboration} {HPS}),\ }\bibfield
  {title} {\bibinfo {title} {{Search for a dark photon in electroproduced
  $e^{+}e^{-}$ pairs with the Heavy Photon Search experiment at JLab}},\ }\href
  {https://doi.org/10.1103/PhysRevD.98.091101} {\bibfield  {journal} {\bibinfo
  {journal} {Phys. Rev. D}\ }\textbf {\bibinfo {volume} {98}},\ \bibinfo
  {pages} {091101} (\bibinfo {year} {2018})},\ \Eprint
  {https://arxiv.org/abs/1807.11530} {arXiv:1807.11530 [hep-ex]} \BibitemShut
  {NoStop}%
\bibitem [{\citenamefont {Bryngemark}\ \emph {et~al.}(2021)\citenamefont
  {Bryngemark} \emph {et~al.}}]{LDCS}%
  \BibitemOpen
  \bibfield  {author} {\bibinfo {author} {\bibfnamefont {L.~K.}\ \bibnamefont
  {Bryngemark}} \emph {et~al.},\ }\bibfield  {title} {\bibinfo {title}
  {{Building a Distributed Computing System for LDMX - Challenges of creating
  and operating a lightweight e-infrastructure for small-to-medium size
  accelerator experiments}},\ }\href
  {https://doi.org/10.1051/epjconf/202125102038} {\bibfield  {journal}
  {\bibinfo  {journal} {EPJ Web Conf.}\ }\textbf {\bibinfo {volume} {251}},\
  \bibinfo {pages} {02038} (\bibinfo {year} {2021})},\ \Eprint
  {https://arxiv.org/abs/2105.02977} {arXiv:2105.02977 [hep-ex]} \BibitemShut
  {NoStop}%
\end{thebibliography}%

\newpage

\begin{appendices}

\section{Energy Dependence of BDT Variables}
\FloatBarrier

\label{app:energyComparison}
\label{app:EnergyComparison}

\begin{figure}[h]
    \centering
    \smaller
    \def\svgwidth{0.95\columnwidth}
    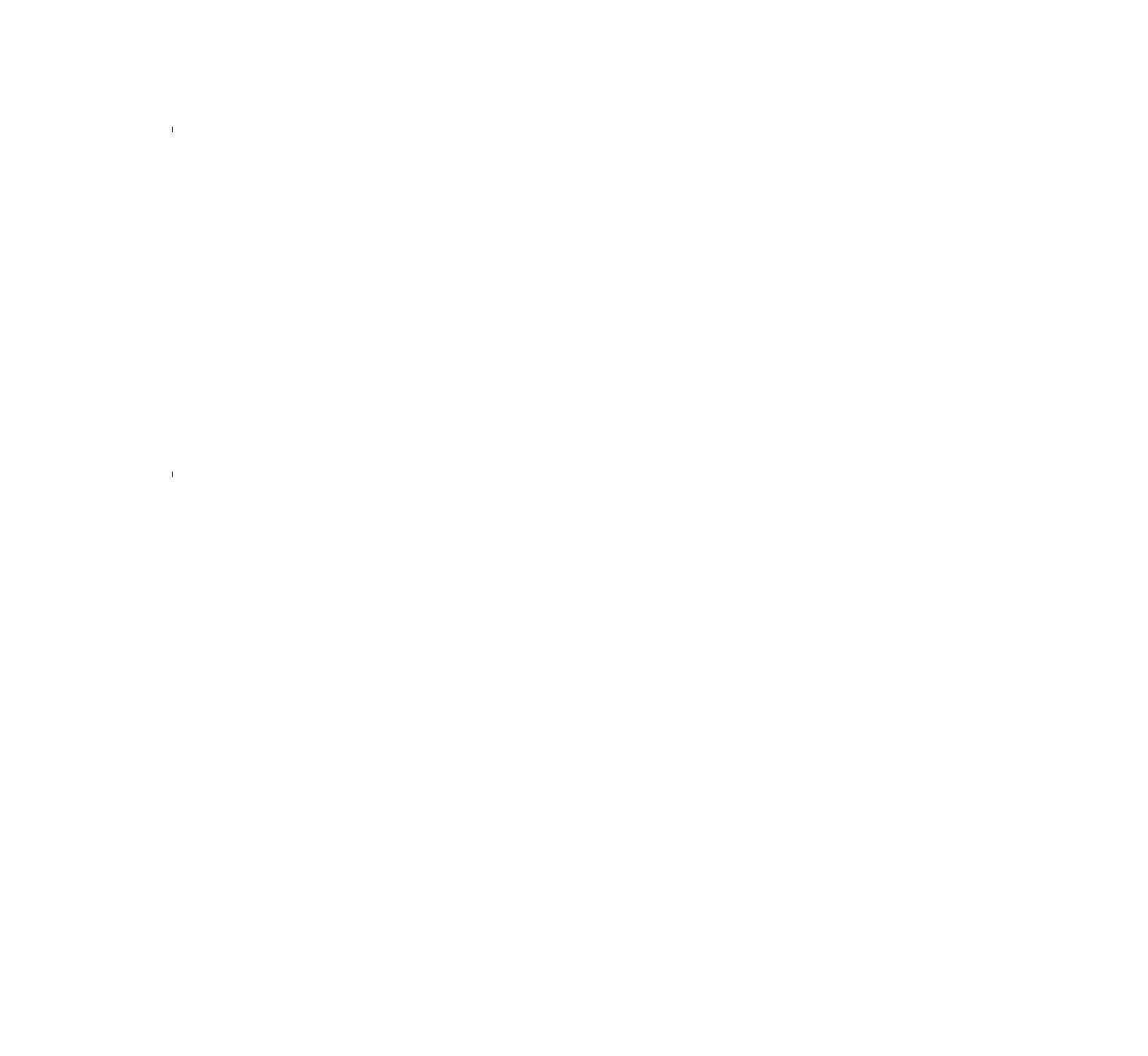
    \label{fig:quadEnergyComparison2}
    \caption{ Distributions of selected BDT variables, comparing signal events with $m_{A'} = 1$ MeV and ECal photo-nuclear events, at both 4 GeV and 8 GeV beam energy. a) Total energy in the last 14 ECal layers. b) Total energy in cells more than four containment radii away from the projected recoil electron and bremsstrahlung photon trajectories. c) Total energy in cells within one containment radii around the projected bremsstrahlung photon trajectory. d) Total energy in cells within one containment radii around the projected recoil electron trajectory.  The rightmost bins are overflow bins.}
\end{figure}

\begin{figure}
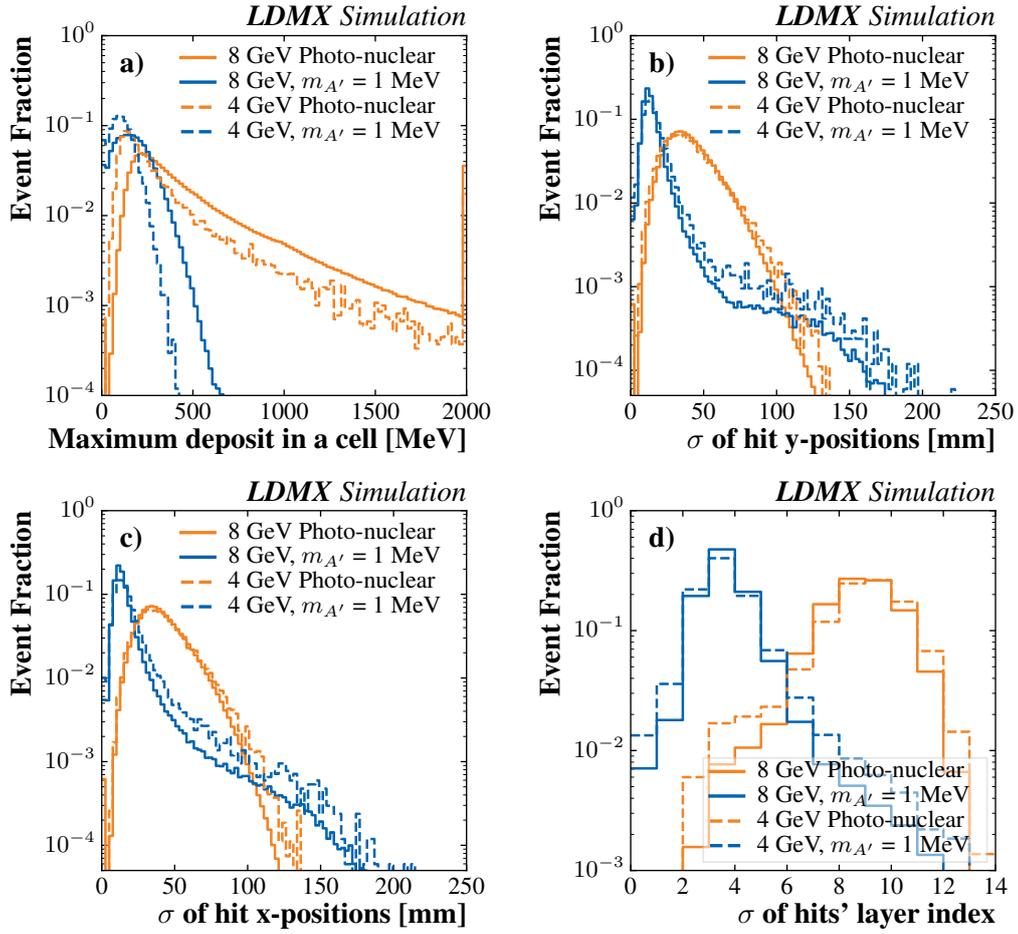

    \centering
    \smaller
    \def\svgwidth{0.95\columnwidth}
    
    \label{fig:quadEnergyComparison3}
    \caption{Distributions of selected BDT variables, comparing signal events with $m_{A'} = 1$ MeV and ECal photo-nuclear events, at both 4 GeV and 8 GeV beam energy. a) The largest energy deposit made to any one ECal cell. b) Standard deviation of all ECal hit's y-position. c) Standard deviation of all ECal hit's x-position. d) Standard deviation of all ECal hit's layer index.  The rightmost bins are overflow bins.}
\end{figure}

\FloatBarrier

\clearpage
\section{BDT Variables at 8 GeV}
\label{app:BDTVariables}

\begin{figure}[h]
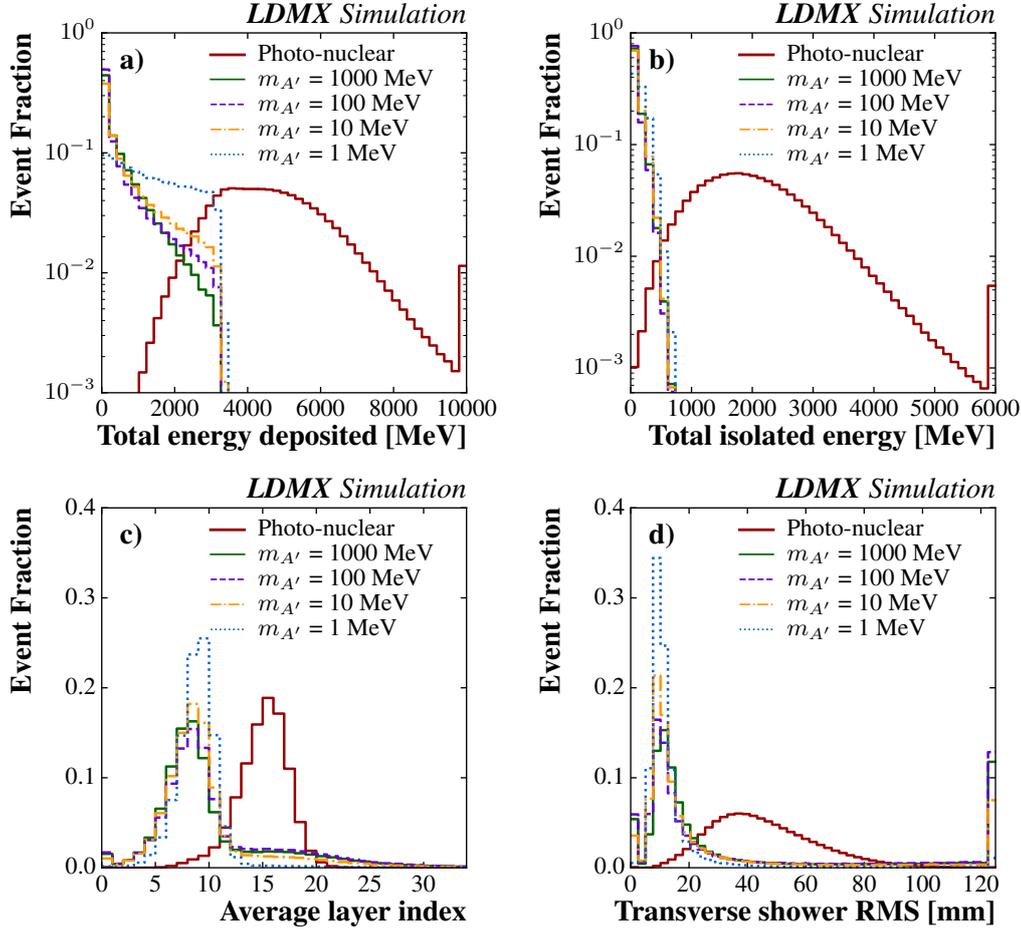

    \centering
    \smaller
    \def\svgwidth{0.95\columnwidth}
    
    \label{fig:quad1}
    \caption{ Distributions of selected BDT variables at 8 GeV, comparing signal events of various mediator masses and ECal photo-nuclear events.
    a) The sum of the energy in each ECal cell. The effect of the trigger is seen in the steep drop in signal events around 3160 MeV, while photo-nuclear events often deposit energy deeper in the ECal. 
    b) Sum of energy in ECal cells for which neighbouring cells in the same layer have no activity.
    c) Average layer index of ECal hits.
    d) Energy-weighted RMS of ECal hit's transverse distance from the energy centroid. The rightmost bins are overflow bins.}
\end{figure}

\begin{figure}
    \centering
    \smaller
    \def\svgwidth{0.95\columnwidth}
    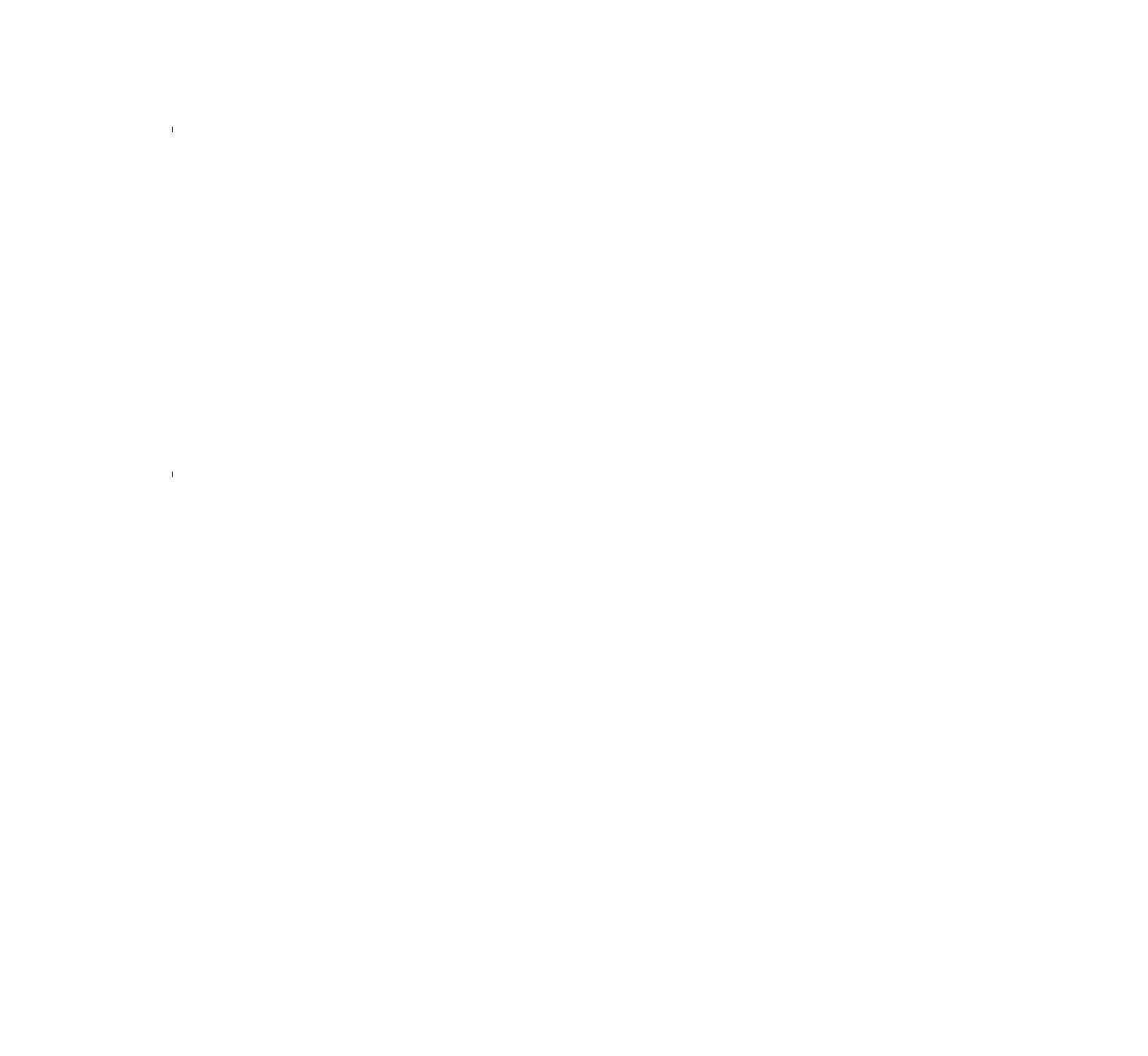
    
    \caption{
    Distributions of selected BDT variables at 8 GeV, comparing signal events of various mediator masses and ECal photo-nuclear events. a) The largest deposit made to any one ECal cell. b) Standard deviation of ECal hit's y-position. c) Standard deviation of ECal hit's x-position. d) Standard deviation of ECal hit's layer index. 
    The rightmost bins are overflow bins.}
    \label{fig:quad3}
\end{figure}

\clearpage
\section{BDT and HCal Complementarity}
\FloatBarrier
\label{app:2dplot}

\begin{figure}[h]
    \centering
    \smaller
    \def\svgwidth{0.93\columnwidth}
    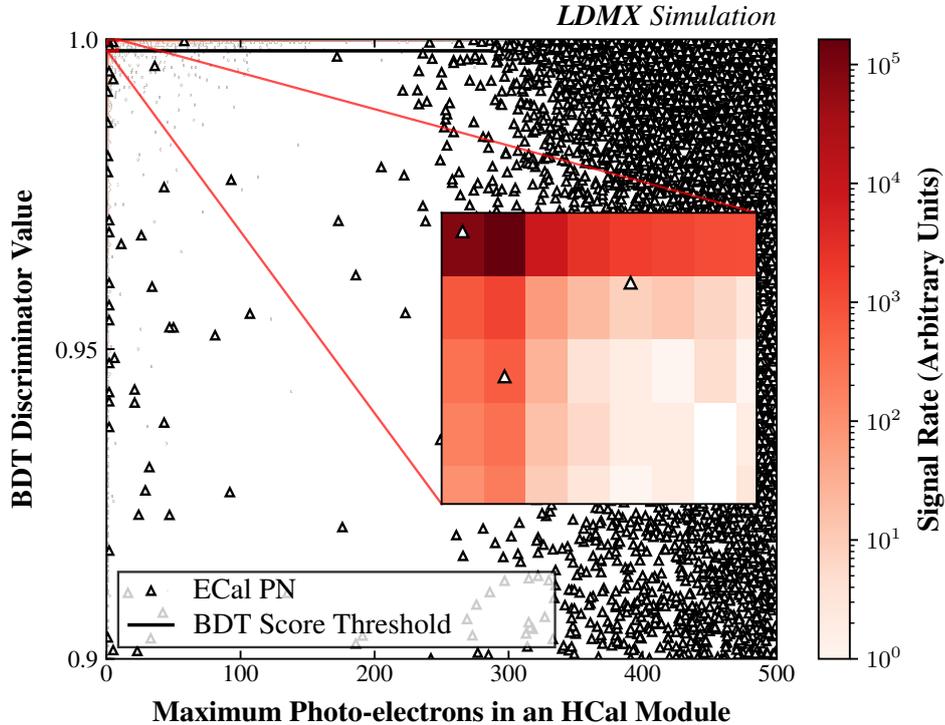
    \caption{Distribution of BDT discriminator values and maximum photo-electrons in the HCal, in ECal photo-nuclear events.
    The inset shows the signal region.
    The rate of $m_{A'} = 1000$ MeV signal events is shown underneath.}
    \label{fig:PNPE}
\end{figure}

\FloatBarrier
\section{Simulation Cutflows}
\label{app:cutflow}

\begin{table}
\centering
\begin{tabular}{l|c|c|c|c}
\hline \hline
 \multicolumn{1}{l|}{} &  \multicolumn{2}{l|}{{\centering \bf \hfil Photo-nuclear}}    &  \multicolumn{2}{l}{\bf \centering \hfil  Muon conversion}     \\ \cline{2-5}
  &  {\bf \centering Target-area} & {\bf \centering ECal}  & {\bf Target-area} & {\bf ECal}   \\ \hline
 EoT Equivalent & $8.99\times10^{14}$ & $1.98 \times 10^{14}$ & $9.45\times10^{15}$ & $2.40\times10^{15}$  \\ \hline \hline
 Trigger (front ECal energy $ < 3160$ MeV) & $3.40\times10^8$  & $4.39\times10^8$ &  $1.12\times10^9$ & $9.73\times10^8$  \\ 
 Total ECal energy $ < 3160$ MeV & $1.23\times10^8$ & $7.19\times10^7$ & $8.29\times10^8$ & $7.27\times10^8$ \\
 Single track with p $< 2400$ MeV/c & $1.36\times10^7$  & $6.57\times10^7$  & $2.51\times10^6$  & $6.82\times10^8$  \\ 
ECal BDT (85\%  eff. $m_{A'} = 1$ MeV ) & $6.76\times10^5$  & $1.03\times 10^5$ & 2.0 & 5.0  \\ 
 HCal max PE $< 8$  & 0 & 2.0  & 0 & 3.0 \\ 
 ECal MIP tracks = 0 & 0 & 0 & 0 & 0 \\ \hline
\end{tabular}
\caption{Number of simulated background events surviving the different selection stages. The top row gives the number of EoT corresponding to the generated sample size.}
\label{tab:cutflowRaw}
\end{table}

\end{appendices}

\end{document}